\documentclass[a4paper,twocolumn,11pt]{quantumarticle}
\pdfoutput=1

\usepackage[utf8]{inputenc}
\usepackage[english]{babel}
\usepackage{amsmath,amssymb,amsfonts,amsthm,mathtools}
\usepackage{graphicx}
\usepackage{tikz}
\usetikzlibrary{positioning,arrows.meta}
\usepackage{float,placeins}
\usepackage{hyperref}
\usepackage{algorithm,algorithmic}
\usepackage{caption,subcaption}
\usepackage{braket,setspace,booktabs,array}
\usepackage{xcolor}
\usepackage{microtype}

\newcommand{\safeinclude}[2]{%
  \IfFileExists{#1}{\includegraphics[width=#2]{#1}}{%
    \fbox{\parbox[c][3.4cm][c]{0.90\linewidth}{\centering
      Missing final figure: \texttt{\detokenize{#1}}.\\
      Replace this box with the full-run output before submission.}}}}

\begin{document}

\title{Physics-Informed Support Vector Kernels via Green-Function Analogies and Jackson--Chebyshev Spectral Design}

\author{Nan-Hong Kuo}
\affiliation{Department of Physics, National Taiwan University, Taipei, Taiwan}
\email{d91222008@ntu.edu.tw}

\author{Renata Wong}
\affiliation{Department of Artificial Intelligence, Chang Gung University, Taoyuan, Taiwan}
\affiliation{Department of Neurology, Chang Gung Memorial Hospital, Keelung, Taiwan}
\affiliation{Artificial Intelligence Research Center, Chang Gung University, Taoyuan, Taiwan}
\email{renata.wong@cgu.edu.tw}
\orcid{0000-0001-5468-0716}
\thanks{corresponding author}

\maketitle

\begin{abstract}
Kernel selection for regression of physical observables is often heuristic. We investigate a physics-informed strategy in which functional forms and spectral structures associated with Green's functions motivate kernel selection without requiring an exact identification between a machine-learning kernel and a physical propagator. The principal construction is a Jackson-damped Chebyshev kernel inspired by the kernel polynomial method (KPM); its explicit feature map yields a positive-semidefinite Gram matrix by construction and provides an inspectable spectral prior for structured observables. We evaluate standard and custom SVR models on copper-conductivity proxies, local Dirac-like band dispersion, quartic-oscillator energy levels, photonic-crystal transmission, and Fibonacci-chain transmission using repeated nested validation, learning curves, random-forest and multilayer-perceptron baselines, and low-rank Nystr\"om tests where relevant. 
The framework is intended for finite-data regression of precomputed observables while boundary conditions remain part of the physical model that generates those observables.
\end{abstract}

\section{Introduction}
Machine learning is now routinely used to model physical systems when analytical descriptions are unavailable, expensive to evaluate, or incomplete. In kernel methods, one of the central modeling choices is the kernel itself: it determines the geometry of the feature space in which the regression problem is solved. Standard choices such as linear, polynomial, and radial-basis-function (RBF) kernels are effective in many settings, but their selection is often empirical rather than tied to the structure of the underlying physical problem \cite{Cortes1995,Vapnik1999,Scholkopf2002,Drucker1997}.

Green's functions and propagators provide a complementary language for physical response. They encode how an operator, together with its domain and boundary conditions, organizes spectral modes and mediates the response to a source or perturbation \cite{Feynman1965,Sakurai2014,Mahan1990}. Connections between Green-function constructions and kernel methods are not new: discrete Green functions have been related to discrete kernel functions \cite{Deeter1974}, heat kernels to Green-function bounds \cite{Davies1987}, radial-basis and spline kernels to differential-operator Green functions \cite{Fasshauer2011}, and fermionic correlation kernels to single-particle Green functions \cite{Dean2021}. Related machine-learning work has also learned Green-function-type solution operators for differential equations \cite{Gin2021,Li2020}.

These connections motivate the question studied here: can functional and spectral structure associated with a physical response be used as a practical prior for kernel selection and construction? We use the Green-function connection at this restricted level. In a special case, the free Euclidean propagator has exactly the Gaussian spatial dependence of an RBF kernel. More generally, Mercer kernels and Green functions both admit spectral representations, although their coefficients and mathematical roles are different. For observables with strongly structured spectral dependence, we construct a Jackson-damped Chebyshev feature map inspired by KPM \cite{Silver1994,Weisse2006,JacobiKPM2025}. The resulting kernel is positive semidefinite by construction because it is an explicit inner product.

Contemporary scientific machine learning also includes physics-informed neural networks (PINNs) and neural-operator architectures such as DeepONet and Fourier neural operators (FNOs) \cite{Raissi2019,Lu2021,LiFNO2021}. These methods are especially natural for solving differential equations over a continuous domain or learning maps between functions. The present setting is complementary: we consider finite-data regression of already generated scalar or low-dimensional observables from a modest number of descriptors. A more detailed methodological comparison, including the specific role of SVR and the empirical diagnostics used here, is given in Sec.~\ref{sec:svr_positioning}.

\subsection*{Contributions}
The main contributions are:
\begin{enumerate}
\item We formulate a controlled functional and spectral analogy between positive-semidefinite kernels and Green-function representations, while keeping the regression Gram matrix distinct from a physical resolvent.
\item We construct a KPM-inspired Jackson--Chebyshev kernel with an explicit feature map, making positive semidefiniteness and the spectral weighting directly inspectable.
\item We evaluate the resulting inductive bias with repeated nested validation, learning curves, random-forest and multilayer-perceptron baselines, kernel-alignment diagnostics, and support-vector statistics across representative physical-regression problems.
\item For the structured photonic and Fibonacci transmission problems, we add a Nystr\"om low-rank representation and measure the tradeoff among landmark rank, Gram-matrix approximation error, predictive accuracy, runtime, and memory.
\end{enumerate}
\section{SVM kernel methods, propagators, and Green's functions}\label{sec:svm-green}
For scalar regression, an $\epsilon$-SVR predictor has the representer form
\begin{equation}
 f(x)=\sum_{i=1}^{n}(\alpha_i-\alpha_i^*)K(x_i,x)+b,
\end{equation}
where the coefficients are obtained from a convex dual problem when $K$ is positive semidefinite \cite{Drucker1997,Vapnik1998}. The kernel trick replaces an ordinary inner product with
\begin{equation}
K(x,x')=\langle\phi(x),\phi(x')\rangle_{\mathcal H}.
\end{equation}
For a finite sample $\{x_i\}_{i=1}^n$, the corresponding \emph{Gram matrix} is $K_{ij}=K(x_i,x_j)$. A Mercer kernel also admits, under the usual assumptions, a positive-semidefinite spectral representation
\begin{equation}
K(x,x')=\sum_j\lambda_j\phi_j(x)\phi_j(x'),\qquad \lambda_j\ge0
\end{equation}
\cite{Mercer1909,KimeldorfWahba1971,Genton2001}. Common choices include the linear, polynomial, RBF, and sigmoid kernels. The sigmoid form may be indefinite outside restricted parameter ranges \cite{SigmoidSMO,SupportVectorIndefiniteKrein}.

A quantum propagator is the matrix element of the time-evolution operator,
\begin{equation}\label{eq:propagator}
U(x',t;x,0)=\langle x'|e^{-iHt/\hbar}|x\rangle,
\end{equation}
while a retarded Green function is the resolvent
\begin{equation}\label{eq:retarded_resolvent}
G^R(E)=(E-H+i0^+)^{-1}.
\end{equation}
The two are connected exactly by the one-sided Fourier--Laplace transform
\begin{equation}\label{eq:resolvent_propagator_transform}
G^R(E)=-\frac{i}{\hbar}\int_0^\infty
 e^{i(E+i0^+)t/\hbar}e^{-iHt/\hbar}\,dt.
\end{equation}
The sign and prefactor in Eq.~\eqref{eq:resolvent_propagator_transform} can be checked on an energy eigenstate $H|j\rangle=E_j|j\rangle$. For a positive infinitesimal $\eta$,
\begin{equation}\label{eq:resolvent_scalar_check}
-\frac{i}{\hbar}\int_0^\infty
 e^{[i(E-E_j)-\eta]t/\hbar}\,dt
 =\frac{1}{E-E_j+i\eta},
\end{equation}
and the retarded prescription follows as $\eta\to0^+$. Thus Eq.~\eqref{eq:resolvent_propagator_transform} is the operator form of Eq.~\eqref{eq:retarded_resolvent}, rather than an additional modeling assumption. For a self-adjoint Hamiltonian the corresponding position-space spectral expansion is
\begin{equation}\label{eq:green_spectral}
G^R(x,x';E)=\sum_j\frac{\psi_j(x)\psi_j^*(x')}{E-E_j+i0^+}.
\end{equation}
The resolvent coefficients are generally complex and sign-indefinite, unlike the nonnegative Mercer eigenvalues. Equations~\eqref{eq:propagator}--\eqref{eq:green_spectral} therefore supply an exact physical relation between propagation and resolvent response, but in this manuscript they are used only to motivate the subsequent functional and spectral analogies; the benchmark kernels are not all obtained by pointwise matching to $G^R$.

\subsection{Gaussian special case}
At fixed imaginary time $\tau>0$, the free Euclidean propagator is
\begin{equation}
G_E(x,x';\tau)=\left(\frac{m}{2\pi\hbar\tau}\right)^{d/2}
\exp\left[-\frac{m\|x-x'\|^2}{2\hbar\tau}\right].
\end{equation}
After removing its positive normalization factor, its spatial dependence is exactly an RBF form with $\gamma=m/(2\hbar\tau)$. This is a special functional identity and does not make an arbitrary RBF-SVR predictor a physical propagator.

\subsection{Operator-level spectral analogy}
The operator analogy is most useful at the level of basis structure. A Gram matrix may be written $K=V\Lambda V^\top$, with nonnegative eigenvalues for a valid kernel, whereas a Green function is the resolvent of a specified operator and its domain. Both organize response through rank-one modes, but their coefficients and mathematical roles differ. In particular, an SVM kernel $K(x,x')=\langle\phi(x),\phi(x')\rangle_{\mathcal H}$ is an inner product in feature space, whereas $G^R(x,x';E)=\langle x|G^R(E)|x'\rangle$ is a matrix element of an operator inserted between states. The latter is not, in general, a positive feature-space inner product. Likewise, the operator inverse $[G^R(E)]^{-1}=E-H+i0^+$ should not be confused with either the pointwise reciprocal $1/K(x,x')$ or the inverse of a finite Gram matrix. The SVR dual coefficients are optimization variables rather than physical source amplitudes, and a label-weighted Gram matrix is not a Hamiltonian resolvent. The analogy therefore motivates choosing an explicit spectral basis; it does not convert the regression problem into the underlying quantum boundary-value problem.

\begin{figure*}[htb]
\centering
\begin{tikzpicture}[>={Stealth[length=2.2mm]},thick,font=\small,
box/.style={draw,rounded corners,align=center,minimum height=1.0cm,inner xsep=5pt,inner ysep=4pt}]
\node[box,text width=3.0cm] (data) {data and target\\observable};
\node[box,text width=3.2cm,right=0.70cm of data] (kernel) {PSD kernel\\$K=\Phi\Phi^{\top}$};
\node[box,text width=2.8cm,right=0.70cm of kernel] (svr) {convex SVR\\predictor};
\draw[->] (data.east)--(kernel.west); \draw[->] (kernel.east)--(svr.west);
\node[box,text width=4.7cm,below=1.45cm of kernel] (physics) {physical operator, domain,\\and boundary conditions};
\node[box,text width=2.9cm,right=0.70cm of physics] (green) {Green's function\\/ observable};
\draw[->] (physics.east)--(green.west);
\draw[dashed,<->,shorten <=2pt,shorten >=2pt] (kernel.south)--node[midway,right=3pt,align=left,fill=white,inner sep=1.5pt]{functional and\\spectral analogy}(physics.north);
\end{tikzpicture}
\caption{Restricted scope of the analogy. The physical model fixes the operator domain and boundary conditions; the regression kernel supplies an inductive bias for interpolation of the resulting observable.}
\label{fig:analogy}
\end{figure*}

\subsection{Scope of the analogy and boundary conditions}\label{sec:scope}
The similarity between a kernel expansion and a Green-function representation does not make the two optimization problems identical. SVR dual coefficients are learned optimization variables, whereas the source term in a physical boundary-value problem is part of the model specification. The distinction becomes especially important for boundary conditions, because a Green function is the inverse of a particular operator together with its domain, not merely of a differential expression.

A simple one-dimensional example illustrates the point. Consider the differential expression $-d^2/dx^2$ on $0<x<L$. With Dirichlet conditions, $u(0)=u(L)=0$, the eigenmodes are proportional to $\sin(n\pi x/L)$ with $n\ge1$. With Neumann conditions, $u'(0)=u'(L)=0$, the eigenmodes are proportional to $\cos(n\pi x/L)$ and include a zero mode. Thus the same differential expression produces a different spectrum, and therefore a different Green function, when the operator domain is changed by the boundary conditions.

For comparison, suppose two operators $A$ and $B$ act on a common domain and define resolvents $R_A(z)=(z-A)^{-1}$ and $R_B(z)=(z-B)^{-1}$. Their second resolvent identity is
\begin{equation}\label{eq:second_resolvent}
R_A(z)-R_B(z)=R_A(z)(A-B)R_B(z).
\end{equation}
If $A-B$ is bounded and $z$ belongs to both resolvent sets, submultiplicativity of the operator norm gives
\begin{equation}\label{eq:resolvent_bound}
\|R_A(z)-R_B(z)\|\le
\|R_A(z)\|\,\|A-B\|\,\|R_B(z)\|.
\end{equation}
Equation~\eqref{eq:resolvent_bound} is a perturbation estimate for operators satisfying these common-domain assumptions. A change of boundary condition can change the operator domain itself, so it need not be representable as a small bounded perturbation $A-B$. Consequently, Eqs.~\eqref{eq:second_resolvent}--\eqref{eq:resolvent_bound} are not a universal error bound for omitted boundary conditions in a regression model.

The remedy used here is therefore physical rather than statistical: boundary conditions are imposed in the data generator before regression. The photonic transfer-matrix calculation includes the incident medium, finite stack, and exit medium, while the Fibonacci transport calculation includes retarded self-energies for semi-infinite leads. The regression kernel is applied only to the resulting boundary-conditioned observable; it is not claimed to reconstruct or compensate for missing boundary physics.

Boundary information can, in principle, also be built into the kernel itself. If $\{\varphi_j^{(B)}\}$ is a basis adapted to a specified boundary condition $B$, with each basis function satisfying that condition, then any nonnegative sequence $\mu_j$ defines
\begin{equation}\label{eq:boundary_adapted_kernel}
K_B(x,x')=\sum_j \mu_j\,\varphi_j^{(B)}(x)\varphi_j^{(B)*}(x'),\qquad \mu_j\ge0,
\end{equation}
which is positive semidefinite and explicitly boundary dependent. A simpler data-driven alternative is to include boundary descriptors $b$ in the input and learn with a kernel $K((x,b),(x',b'))$. These are natural extensions when several boundary conditions must be learned within one surrogate model. They are not implemented in the present benchmarks, whose purpose is to keep the physical boundary-value problem in the simulator and to study the regression prior separately.

\subsection{Methodological role of finite-data SVR}\label{sec:svr_positioning}
The use of SVR in this work is tied to the structure of the learning problem rather than to a claim that SVMs are universally preferable to modern neural methods. PINNs are designed to enforce differential equations and boundary or initial conditions through the training objective, while DeepONet and FNO-type models learn maps between input and output functions. By contrast, the present tasks take a finite set of precomputed scalar or low-dimensional observables and regress them from a small number of descriptors. The distinction is useful because the methods address different learning objects rather than a single interchangeable benchmark.

\begin{table*}[htb]
\centering
\scriptsize
\caption{Methodological positioning of the learning problems considered here. The table distinguishes methods by the object that is learned rather than treating all scientific-ML architectures as interchangeable baselines.}
\label{tab:method_positioning}
\begin{tabular}{@{}lllll@{}}
\toprule
\parbox[t]{2.2cm}{Method} &
\parbox[t]{3.0cm}{Primary learning object} &
\parbox[t]{3.7cm}{Where physical structure enters} &
\parbox[t]{3.2cm}{Optimization / representation} &
\parbox[t]{2.6cm}{Role in this work} \\
\midrule
\parbox[t]{2.2cm}{Explicit-kernel SVR} &
\parbox[t]{3.0cm}{Scalar or low-dimensional observable from descriptors} &
\parbox[t]{3.7cm}{Physical data generator and chosen kernel/feature map} &
\parbox[t]{3.2cm}{Convex for a fixed PSD kernel; explicit similarities or features} &
\parbox[t]{2.6cm}{Main method under study} \\
\addlinespace
\parbox[t]{2.2cm}{PINN} &
\parbox[t]{3.0cm}{Field solving an ODE /PDE} &
\parbox[t]{3.7cm}{Differential-equation residual plus initial/boundary constraints in the loss} &
\parbox[t]{3.2cm}{Nonlinear neural network training} &
\parbox[t]{2.6cm}{Complementary problem class} \\
\addlinespace
\parbox[t]{2.2cm}{DeepONet/FNO} &
\parbox[t]{3.0cm}{Function-to-function operator map} &
\parbox[t]{3.7cm}{Training pairs of input / output functions and architectural / operator priors} &
\parbox[t]{3.2cm}{Nonlinear operator learning model} &
\parbox[t]{2.6cm}{Complementary problem class} \\
\addlinespace
\parbox[t]{2.2cm}{Random forest /MLP} &
\parbox[t]{3.0cm}{Generic supervised input--output map} &
\parbox[t]{3.7cm}{Primarily through descriptors and training data} &
\parbox[t]{3.2cm}{Tree ensemble / nonlinear neural network} &
\parbox[t]{2.6cm}{Direct task-matched baselines} \\
\bottomrule
\end{tabular}
\end{table*}

For a fixed positive-semidefinite kernel, SVR provides three features that are useful in this finite-data setting. First, the optimization problem is convex, so variation in the learned predictor is not confounded by local minima of a deep-network training objective. Second, the Jackson--Chebyshev feature map makes the basis dimension and damping coefficients explicit. Third, kernel-target alignment, support-vector fraction, learning curves, and Nystr\"om rank sweeps provide direct diagnostics of how the chosen inductive bias interacts with the data. These latter two properties arise from the kernel representation rather than from SVR specifically; the same kernel could also be used with kernel ridge regression. Here, SVR is used as a regularized regression estimator with an $\epsilon$-insensitive loss, providing a controlled framework for evaluating the behavior of the proposed kernel under finite-data conditions.

The corresponding evidence is therefore deliberately testable and empirical. The standard-kernel examples ask whether a simple physical cue identifies a parsimonious candidate family without degrading generalization, while the structured-spectrum examples ask whether an explicit spectral basis remains interpretable as its rank, approximation error, predictive error, time, and memory are varied. Random forests and MLPs are the direct task-matched baselines; PINNs and neural operators are discussed to delimit the problem class rather than to create an artificial cross-task accuracy contest. Figures~\ref{fig:copper}--\ref{fig:quasi_scaling} provide the empirical checks used for this positioning.

\section{KPM-Inspired Kernel Construction via Jackson-Damped Chebyshev Expansion}\label{sec:kpm}
The methodological core of this work is a custom kernel construction motivated by the spectral logic of the kernel polynomial method (KPM). KPM uses polynomial expansions to approximate densities of states, correlation functions, and other operator-dependent spectral quantities \cite{Silver1994,Weisse2006,JacobiKPM2025}. We adapt two ingredients of that construction to define an explicit feature representation for regression: Chebyshev basis functions and Jackson damping. The goal is not to reproduce a physical Green function, but to encode a controlled spectral prior in a positive-semidefinite kernel.

Indefinite similarities arise in some kernel formulations, including sigmoid kernels outside restricted parameter regimes \cite{SigmoidSMO,SupportVectorIndefiniteKrein}. Spectral correction, pseudo-Euclidean or Krein-space embedding, and difference-of-convex optimization remain appropriate tools when a genuinely indefinite similarity must be used \cite{TrainingSVMIndefinite,SVMIndefiniteClassification,SupportVectorIndefinite,SolvingIndefiniteSVM,IndefiniteKernelsPCA}. The Jackson--Chebyshev kernel considered here belongs to a different case: it is defined as an explicit inner product and is therefore positive semidefinite analytically. 

\subsection{Custom kernel design with KPM}\label{sec:kpm_design}
To address the potential mismatch between generic kernels and structured physical spectra, we use a constructive framework based on Chebyshev spectral approximation, Jackson damping, and an explicit feature map. For clarity, the formulation is written for one-dimensional scalar spectral inputs. Possible multidimensional extensions, for example through tensor-product Chebyshev features, separable constructions, or operator-valued spectral expansions, are left for future work.

\paragraph{Rescaling.}
The scalar input $x$ is mapped affinely to the Chebyshev interval,
\begin{equation}\label{eq:cheb_rescale}
\tilde x\in[-1+\delta,1-\delta],\qquad \delta>0,
\end{equation}
where the small endpoint offset avoids evaluating numerically at $\pm1$. The Chebyshev polynomials of the first kind are
\begin{equation}\label{eq:cheb_polynomial}
T_j(\tilde x)=\cos\!\left[j\arccos(\tilde x)\right].
\end{equation}

\paragraph{Jackson coefficients.}
For truncation degree $N$, define $\theta_j=j\pi/(N+1)$. The Jackson damping coefficients are
\begin{equation}\label{eq:jackson_coeff}
g_j=\frac{(N-j+1)\cos\theta_j+\sin\theta_j\cot\theta_1}{N+1},
\end{equation}
for $j=0,\ldots,N$. The coefficients are nonnegative and progressively damp the high-order components of the truncated expansion.

\paragraph{Feature mapping.}
For each rescaled input $\tilde x_i$, define
\begin{equation}\label{eq:jackson_feature}
\Phi_N(\tilde x_i)=
[\sqrt{g_0}T_0(\tilde x_i),\ldots,\sqrt{g_N}T_N(\tilde x_i)]^\top.
\end{equation}
The square-root weighting makes the kernel an ordinary Euclidean inner product in this explicit feature space.

\paragraph{Kernel construction and positive semidefiniteness.}
The custom kernel is
\begin{equation}\label{eq:jackson}
\begin{aligned}
K_N(x_i,x_j)&=\Phi_N(\tilde x_i)^\top\Phi_N(\tilde x_j)\\
&=\sum_{r=0}^{N}g_rT_r(\tilde x_i)T_r(\tilde x_j).
\end{aligned}
\end{equation}
For $n$ sampled inputs, let $\Phi\in\mathbb{R}^{n\times (N+1)}$ collect the feature vectors by rows. The corresponding Gram matrix is $K=\Phi\Phi^\top$. For every $c\in\mathbb{R}^n$,
\begin{equation}\label{eq:jackson_psd}
c^\top Kc=c^\top\Phi\Phi^\top c
=\|\Phi^\top c\|_2^2\ge 0.
\end{equation}
Equation~\eqref{eq:jackson_psd} establishes positive semidefiniteness directly from the feature construction; no full-matrix spectral clipping is required for this kernel. In finite-precision calculations, the computed Gram matrix may be symmetrized as
\begin{equation}\label{eq:roundoff_symmetrization}
K\leftarrow\frac{1}{2}(K+K^\top),
\end{equation}
and, if needed for numerical conditioning, a small diagonal regularization $\varepsilon I$ may be added. These operations address roundoff and conditioning only and do not alter the analytical kernel definition.

All the above steps are summarized in Algorithm~\ref{alg:kpm_psd}.

\begin{algorithm}[H]
\caption{PSD-by-construction Jackson--Chebyshev kernel for scalar spectral inputs}
\label{alg:kpm_psd}
\begin{algorithmic}[1]
\REQUIRE Training inputs $X=\{x_1,\ldots,x_n\}$, degree $N$, endpoint offset $\delta$
\ENSURE Feature matrix $\Phi$ and PSD Gram matrix $K$
\STATE Fit an affine map from the training range to $[-1+\delta,1-\delta]$
\STATE Compute $g_0,\ldots,g_N$ from Eq.~\eqref{eq:jackson_coeff}
\FOR{each input $x_i$}
\STATE Set $\Phi_{ir}=\sqrt{g_r}T_r(\tilde x_i)$ for $r=0,\ldots,N$
\ENDFOR
\STATE Set $K=\Phi\Phi^\top$
\STATE Optionally apply Eq.~\eqref{eq:roundoff_symmetrization} and a small diagonal regularization for numerical conditioning
\RETURN $\Phi,K$
\end{algorithmic}
\end{algorithm}

\subsection{Numerical scope and scalability}\label{sec:kpm_scalability}
The numerical procedure follows the analytical construction above. Because Eq.~\eqref{eq:jackson_psd} proves that the Gram matrix is PSD, an eigendecomposition of the full $n\times n$ Gram matrix is not part of the PSD-enforcement procedure. Finite-precision symmetrization or diagonal regularization is only a numerical safeguard; eigenvalue clipping would be a separate matrix modification.

Let $X=\{x_i\}_{i=1}^{n}$ denote the training inputs, where $n$ is the number of training samples. For Chebyshev truncation degree $N$, the explicit Jackson--Chebyshev feature vector has dimension
\begin{equation}\label{eq:feature_dimension}
D=N+1.
\end{equation}
Evaluating the explicit features for all training samples requires $O(nD)$ operations. If the full Gram matrix
\begin{equation}\label{eq:full_gram}
K(X,X)=\Phi\Phi^\top
\end{equation}
is formed explicitly, its construction requires $O(n^2D)$ operations and $O(n^2)$ memory.

Lanczos \cite{lanczos} and Nystr\"om \cite{nystrom,WilliamsSeeger2001} methods address different large-matrix bottlenecks. Lanczos iteration approximates selected extremal eigenpairs in a Krylov subspace without computing a complete eigendecomposition, whereas a Nystr\"om approximation replaces the full kernel representation by a low-rank representation built from a subset of landmark samples. Since positive semidefiniteness is already guaranteed analytically in Eq.~\eqref{eq:jackson_psd}, Lanczos is not used here as a PSD-repair step. 
For larger data sets, we use a Nystr\"om approximation. Let
\begin{equation}\label{eq:nystrom_landmarks}
L=\{\ell_a\}_{a=1}^{m}\subset X,\qquad m\ll n,
\end{equation}
be a set of $m$ landmark samples. Define $C\in\mathbb{R}^{n\times m}$ and $W\in\mathbb{R}^{m\times m}$ by
\begin{equation}\label{eq:nystrom_CW}
\begin{aligned}
C&=K(X,L), & C_{ia}&=K_N(x_i,\ell_a),\\
W&=K(L,L), & W_{ab}&=K_N(\ell_a,\ell_b).
\end{aligned}
\end{equation}
If $W=U\Lambda U^\top$, its Moore--Penrose pseudoinverse square root is obtained by inverting the positive eigenvalues of $W$ and setting the null-space components to zero. The Nystr\"om feature matrix is then
\begin{equation}\label{eq:nystrom_Z}
Z=CW_+^{-1/2}\in\mathbb{R}^{n\times m},
\end{equation}
and the associated low-rank Gram approximation is
\begin{equation}\label{eq:nystrom_gram}
\widetilde K=ZZ^\top=CW_+^{-1}C^\top.
\end{equation}
Because $\widetilde K$ is written as $ZZ^\top$, it is positive semidefinite by construction. Constructing $C$ from the explicit features costs $O(nmD)$, decomposing the $m\times m$ matrix $W$ costs $O(m^3)$, and forming $Z$ by a dense matrix product costs $O(nm^2)$. The representation requires $O(nm+m^2)$ storage, compared with $O(n^2)$ storage for the full dense Gram matrix.


The Nystr\"om approximation is evaluated empirically by varying the number of landmarks and reporting effective rank, relative Gram-matrix error, minimum eigenvalue, predictive MSE and $R^2$, fitting time, and memory usage. These measurements characterize the rank--accuracy--cost tradeoff rather than assuming it a priori. The explicit matrix $Z$ may be used as an $m$-dimensional representation for subsequent regression.

\subsection{Example: construction for silicon-based photonic crystals}\label{sec:kpm_photonic_example}
To illustrate how the spectral construction is connected to a physical regression problem, consider transmission through a finite one-dimensional Si/SiO$_2$ multilayer. Periodic dielectric contrast produces sharp band edges, oscillatory modes, and structured transmission windows, making this system an appropriate testbed for a Chebyshev representation. The physical transfer-matrix model, including the incident medium, finite layer sequence, and exit medium, is evaluated before regression; the kernel is applied only to the resulting energy-dependent observable.

For photon energy $E$, the wavelength in micrometers is related approximately by $\lambda\simeq1.2398/E$ when $E$ is expressed in electron-volts. Following the empirical model used in the simulation, the dielectric response may be written in the form
\begin{equation}\label{eq:silicon_dielectric}
\epsilon(\lambda)=A+\frac{B}{\lambda^2}+C\lambda^2,
\end{equation}
where $A$, $B$, and $C$ specify the chosen dispersion fit. The photon-energy input is then mapped to the Chebyshev interval using
\begin{equation}\label{eq:photonic_rescale}
\widetilde E=\frac{2E-(E_{\max}+E_{\min})}{E_{\max}-E_{\min}}(1-\delta),
\end{equation}
and Eqs.~\eqref{eq:jackson_feature}--\eqref{eq:jackson} define the corresponding Jackson--Chebyshev representation. Here, we specified the kernel construction. The predictive comparisons, boundary-conditioned simulation details, learning curves, and Nystr\"om rank--accuracy--cost results are reported in Sec.~\ref{sec:crystals}.

\subsection{Interpretability diagnostics}\label{sec:kpm_diagnostics}
For centered $K_c$ and target similarity $Y_c=(yy^\top)_c$, centered kernel--target alignment is
\begin{equation}\label{eq:kernel_alignment}
A(K,y)=\frac{\langle K_c,Y_c\rangle_F}{\|K_c\|_F\|Y_c\|_F}.
\end{equation}
Together with the explicit feature dimension, Jackson weights, and support-vector fraction, this makes the chosen prior inspectable \cite{Cristianini2002}. These diagnostics measure geometric agreement with the target. 

\section{Experimental evaluation}\label{sec:Applications}
Here we present five experiments that deliberately probe different levels of physics-informed guidance. Table~\ref{tab:benchmark_hierarchy} summarizes this hierarchy. 

\begin{table*}[htb]
\centering
\scriptsize
\caption{Hierarchy of physical guidance across the five benchmarks. ``Green-function role'' distinguishes direct use of a physical resolvent from weaker functional or spectral motivation.}
\label{tab:benchmark_hierarchy}
\begin{tabular}{@{}llll@{}}
\toprule
\parbox[t]{2.7cm}{Benchmark} &
\parbox[t]{4.0cm}{Physical structure used} &
\parbox[t]{4.2cm}{Green-function role} &
\parbox[t]{4.4cm}{Kernel implication tested} \\
\midrule
\parbox[t]{2.7cm}{Copper conductivity proxy} &
\parbox[t]{4.0cm}{Activated/nonlinear response in band gap and density; localized smooth interpolation} &
\parbox[t]{4.2cm}{Gaussian Euclidean propagator supplies only a special functional analogy; the conductivity Green function is not reconstructed} &
\parbox[t]{4.4cm}{RBF is a physically plausible candidate, with ranking left to validation} \\
\addlinespace
\parbox[t]{2.7cm}{Local Dirac-like dispersion} &
\parbox[t]{4.0cm}{$E\simeq\hbar v_F q$ in a sufficiently small low-energy window} &
\parbox[t]{4.2cm}{Dirac resolvent is displayed for physical context but is not identified with the ML kernel} &
\parbox[t]{4.4cm}{A low-complexity linear/low-order prior should be competitive locally} \\
\addlinespace
\parbox[t]{2.7cm}{Quartic oscillator} &
\parbox[t]{4.0cm}{Perturbative dependence of $E_n$ on $(n,k,\lambda)$} &
\parbox[t]{4.2cm}{No pointwise Green-function matching is used} &
\parbox[t]{4.4cm}{Polynomial family is a finite-domain approximation; degree is selected empirically} \\
\addlinespace
\parbox[t]{2.7cm}{Photonic crystal} &
\parbox[t]{4.0cm}{Stop bands, interference, and oscillatory spectral response} &
\parbox[t]{4.2cm}{Boundary-conditioned Maxwell/transfer-matrix response is generated physically; no Green-function equality is claimed} &
\parbox[t]{4.4cm}{Jackson--Chebyshev basis tests an explicit structured-spectrum prior} \\
\addlinespace
\parbox[t]{2.7cm}{Fibonacci chain} &
\parbox[t]{4.0cm}{Quasiperiodic resonances and open-system transport} &
\parbox[t]{4.2cm}{Lead-dressed $G^R$ explicitly generates $T(E)$} &
\parbox[t]{4.4cm}{Jackson--Chebyshev basis interpolates the resulting resonance-rich observable} \\
\bottomrule
\end{tabular}
\end{table*}

All model rankings use repeated outer cross-validation with an inner hyperparameter-selection loop; error bars denote 95\% confidence intervals across outer splits. Baselines include standard SVR kernels, a mean predictor, random forest, and MLP. Learning curves assess finite-sample behavior, and the structured transmission tests include Nystr\"om rank/scalability analyses.

\subsection{Copper conductivity proxy}\label{sec:conductivity}
The quantum-mechanical motivation begins with the Kubo--Greenwood structure. In a relaxation-time form,
\begin{equation}
\sigma_{\mu\mu}\propto\sum_k v_\mu(k)^2\tau_k\left(-\frac{\partial f}{\partial\epsilon_k}\right),
\end{equation}
or, after organizing states by energy,
\begin{equation}
\sigma\propto\int N(E)\langle v^2(E)\rangle\tau(E)\left(-\frac{\partial f}{\partial E}\right)dE.
\end{equation}
This expression explains why spectral information near the chemical potential and thermally activated carriers control transport. However, we do not claim to compute electrical conductivity or reconstruct its many-body Green function. To avoid an unstable per-material DOS retrieval in the reproducibility workflow, the implemented target is a lower-cost conductivity proxy: for metallic entries a carrier factor of 1 is used, whereas for gapped entries the factor is $\exp[-E_g/(2k_BT)]$; this is multiplied by density and the base-10 logarithm is regressed.

The input descriptors are therefore band gap and density. The metallic/gapped switch produces a strongly nonlinear, regime-changing target. The RBF kernel is included because its localized Gaussian similarity is compatible with smooth local interpolation and, in the special Euclidean-propagator case of Sec.~\ref{sec:svm-green}, has the same Gaussian form. This is a functional prior, not a statement that the RBF kernel is the conductivity Green function. In the hierarchy of Table~\ref{tab:benchmark_hierarchy}, this example should therefore be read as a functional-form sanity check rather than as a first-principles kernel derivation.

\begin{figure*}[htb]
\centering
\safeinclude{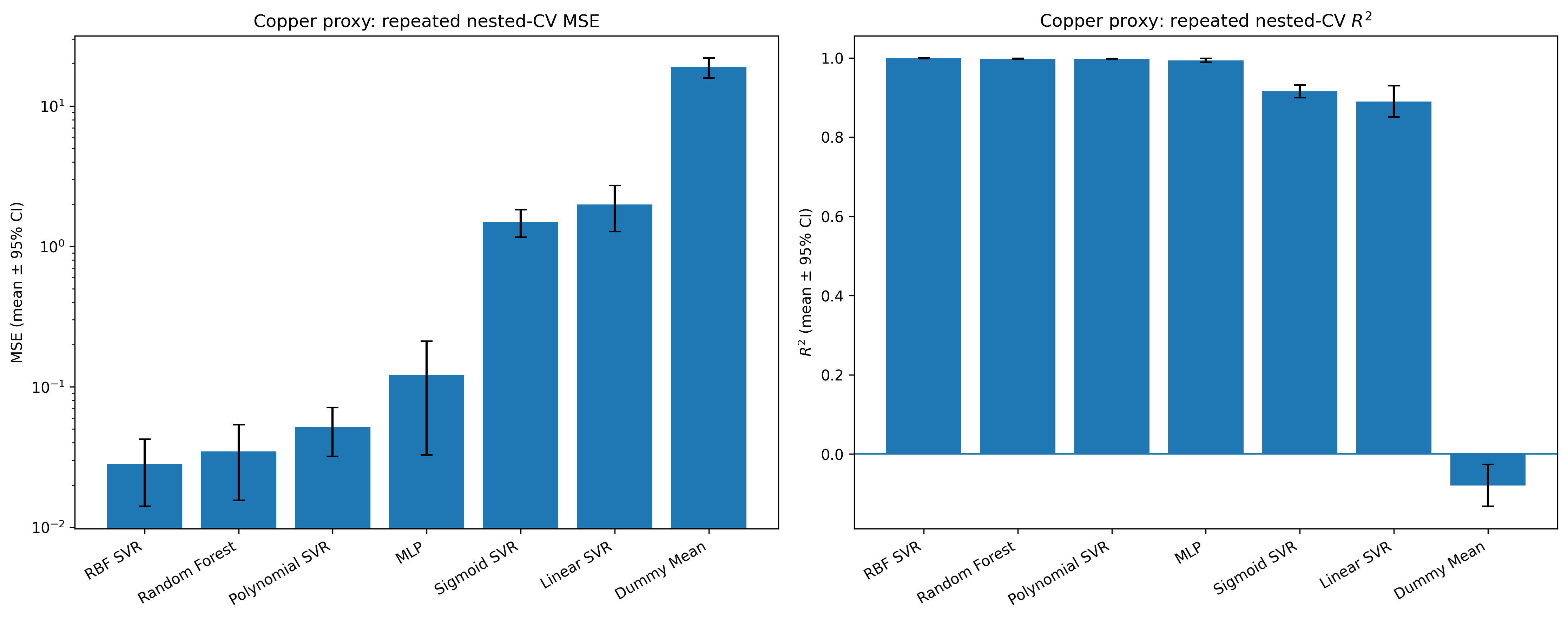}{0.52\textwidth}\hfill
\safeinclude{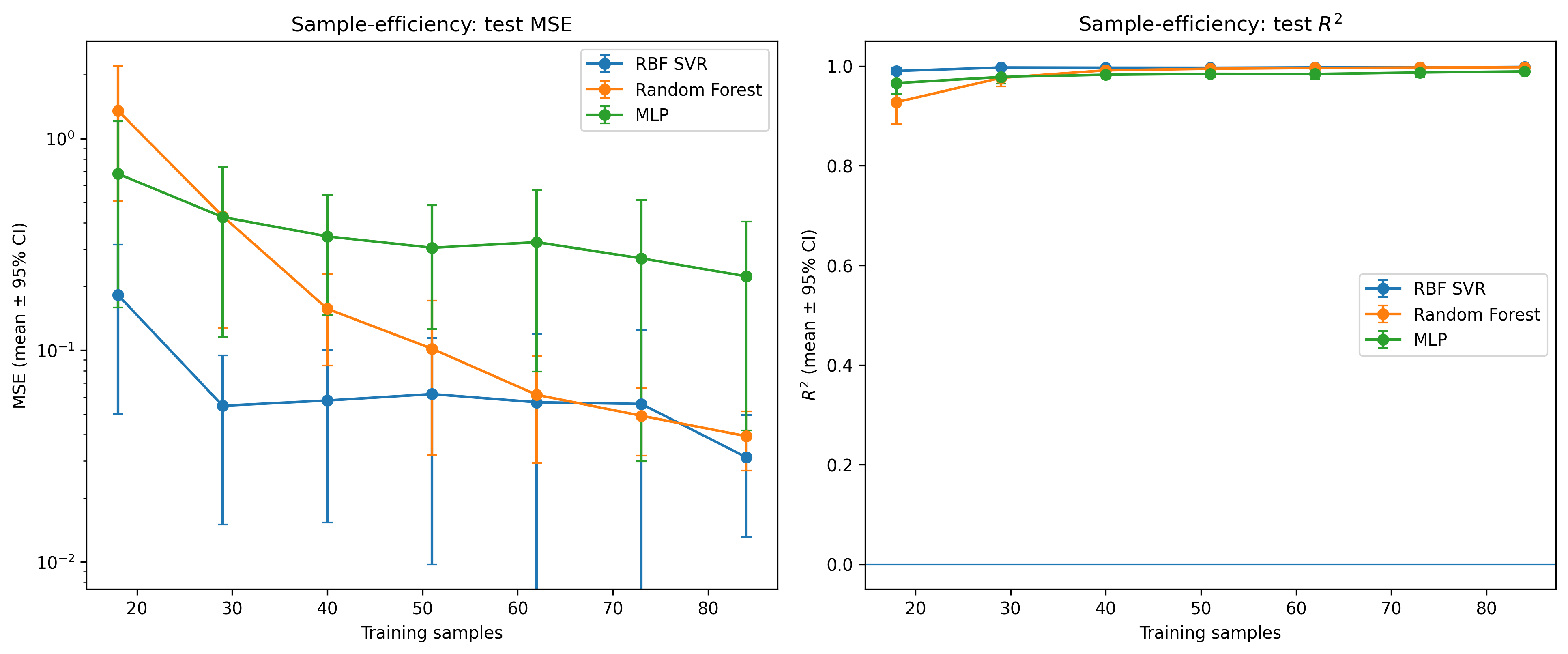}{0.45\textwidth}
\caption{Copper conductivity-proxy benchmark and sample-efficiency analysis. Left: repeated nested-CV mean test MSE (log scale) and $R^2$ with 95\% confidence intervals for the compared SVR and baseline models; lower MSE and higher $R^2$ indicate better generalization. Right: test MSE and $R^2$ versus training-set size for representative leading model families. RBF SVR remains in the leading low-error group across the finite-sample regime, while competitive random-forest performance emphasizes that the Euclidean-propagator analogy supplies a plausible localized-similarity prior rather than a uniqueness theorem. For the copper dataset, the physics-motivated kernel choice is natively the RBF kernel. Consequently, the physics candidate and the standard generic SVR baseline coincide, yielding three distinct model-family benchmarks (RBF SVR, Random Forest, and MLP).}
\label{fig:copper}
\end{figure*}

Figure~\ref{fig:copper} tests whether that simple prior survives contact with the data. The repeated nested-CV panel places RBF SVR in the leading low-error group, while random forest and other flexible regressors remain competitive. The learning curves then ask a different question: whether the low error is retained as the training set is reduced. The resulting behavior supports RBF as a parsimonious finite-data candidate for this proxy, but the competitive baselines also show why the Gaussian analogy alone cannot establish a unique or universally superior kernel.

\subsection{Local Dirac-like band dispersion}\label{sec:band}
Near a Dirac point, the low-energy Hamiltonian has the form \cite{graphene,neto}
\begin{equation}
H_D(\mathbf q)=\hbar v_F\,\boldsymbol{\sigma}\cdot\mathbf q,
\end{equation}
with eigenvalues
\begin{equation}
E_\pm(\mathbf q)=\pm\hbar v_F|\mathbf q|.
\end{equation}
The corresponding retarded Green function is
\begin{equation}
G_D^R(E,\mathbf q)=\left[E+i0^+-H_D(\mathbf q)\right]^{-1}.
\end{equation}
It is a matrix-valued rational resolvent; it is not a linear machine-learning kernel. The physically relevant observation for regression is instead that the local dispersion itself is first order in $q=|\mathbf q|$. A linear kernel is therefore a parsimonious low-bias prior for predicting $E$ from $q$ within a sufficiently small Dirac-like window, while nonlinear kernels test departures from that local approximation. The rationale comes from the observable $E(q)$ generated by the low-energy Hamiltonian, not from taking a pointwise reciprocal of the retarded Green function.

\begin{figure*}[htb]
\centering
\safeinclude{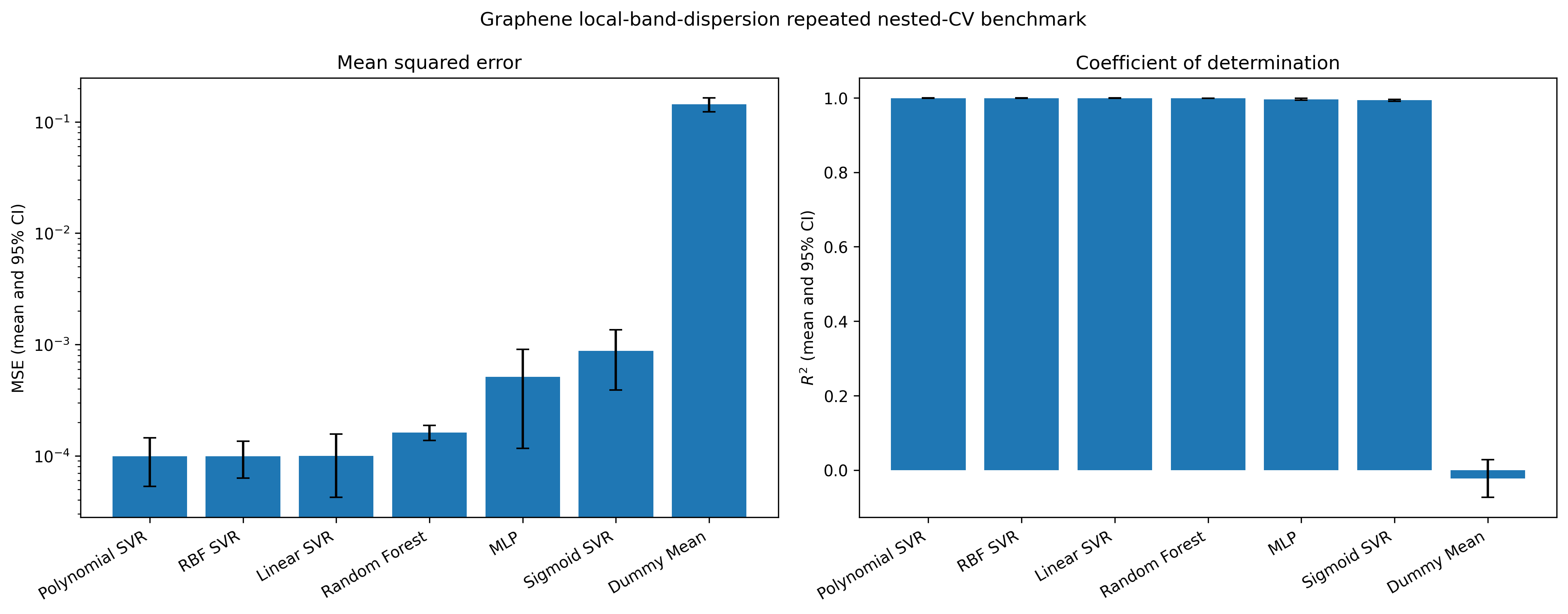}{0.52\textwidth}\hfill
\safeinclude{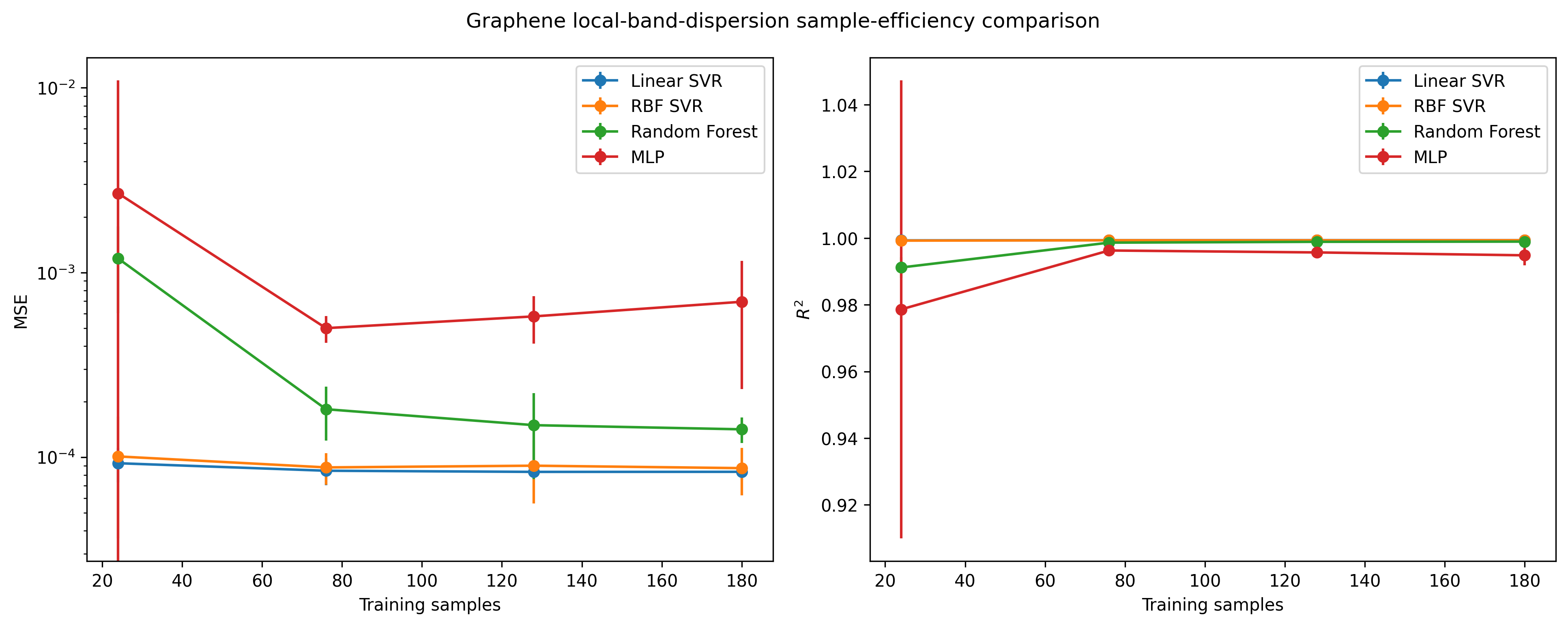}{0.45\textwidth}
\caption{Local Dirac-like band-dispersion benchmark and sample-efficiency analysis. Left: repeated nested-CV MSE (log scale) and $R^2$ for standard SVR kernels and non-kernel baselines. Several smooth low-complexity kernels are nearly degenerate in the small near-Dirac window, consistent with an approximately first-order dispersion rather than with a unique kernel identity. Right: learning curves for representative model families show that the low-error SVR behavior is stable as training size changes. The physical conclusion is therefore a parsimonious low-order prior motivated by $E\propto q$, not an identification of $G_D^R$ with a linear kernel.}
\label{fig:graphene}
\end{figure*}

The Materials Project \cite{materials_project, mp2} extraction uses entry mp-48, identifies a sampled near-Fermi crossing, forms the in-plane distance $q$, and retains conduction-side samples inside a specified energy window. This is a finite high-symmetry-path regression, not a complete Brillouin-zone reconstruction. 
The nested-CV results in Fig.~\ref{fig:graphene} show that several low-complexity smooth kernels perform very similarly in this narrow window, while the learning curves keep linear/RBF SVR at low error as sample size varies. Thus the Hamiltonian identifies a natural simple candidate, but the data do not support a claim that the linear kernel is uniquely optimal.

\subsection{Quartic anharmonic oscillator}\label{sec:anh}
In units $m=\hbar=1$, the numerical Hamiltonian is
\begin{equation}\label{eq:quartic_H}
H=-\frac12\frac{d^2}{dx^2}+\frac12kx^2+\lambda x^4.
\end{equation}
The supervised input is $(n,k,\lambda)$ and the target is the numerically computed eigenvalue $E_n$ \cite{Landau1977}. With $\omega=\sqrt{k}$, first-order Rayleigh--Schr\"odinger perturbation theory gives
\begin{equation}\label{eq:quartic_first_order}
E_n^{(1)}=\sqrt{k}\left(n+\frac12\right)
+\frac{3\lambda}{4k}(2n^2+2n+1).
\end{equation}
Equation~\eqref{eq:quartic_first_order} provides an analytical reference for the weakly anharmonic regime, but it does not prescribe the degree of a polynomial machine-learning kernel. In particular, the leading perturbative correction is quadratic in $n$, while the dependence on the complete input vector remains mixed through $\sqrt{k}$, $1/k$, and $\lambda$; higher perturbative orders add further nonlinear combinations. We therefore treat the polynomial degree as a hyperparameter and select it only through the inner validation loop.

\begin{figure*}[htb]
\centering
\safeinclude{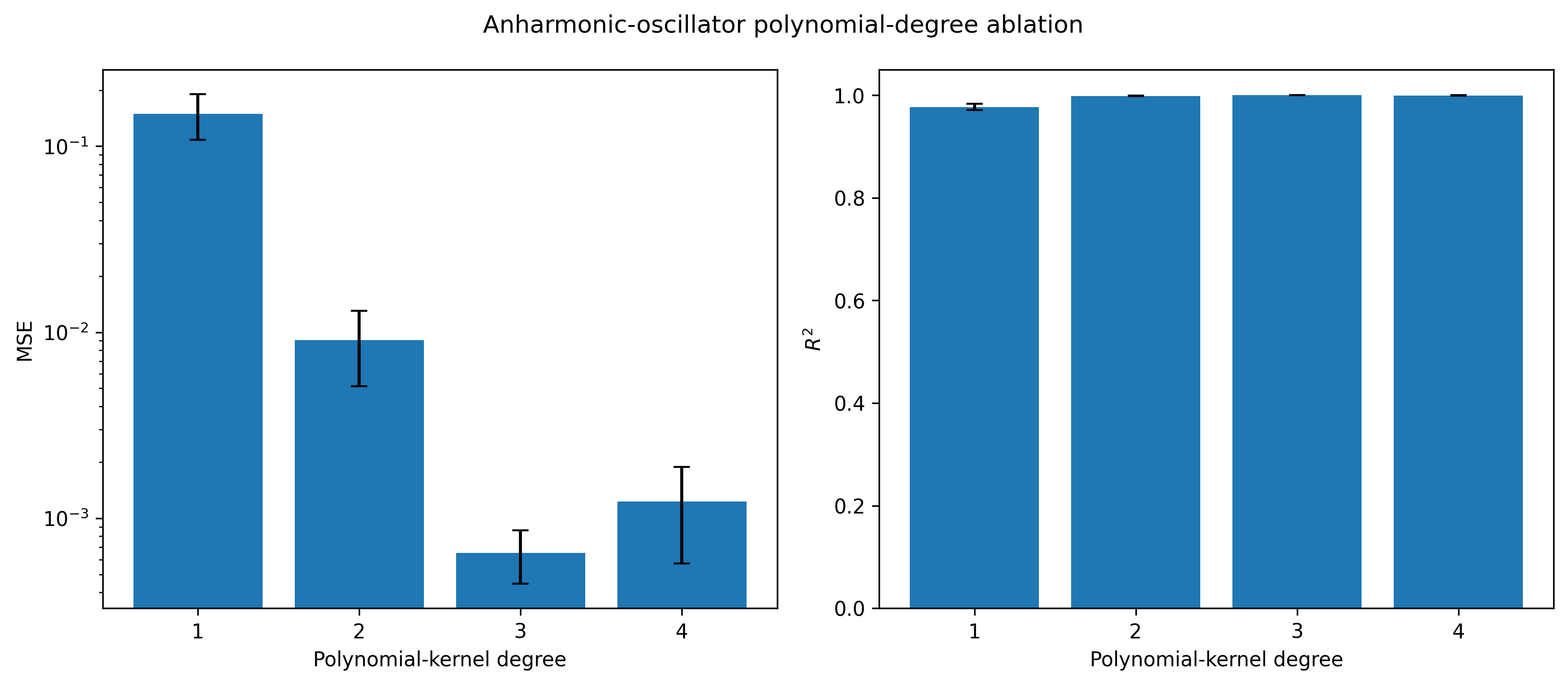}{0.32\textwidth}\hfill
\safeinclude{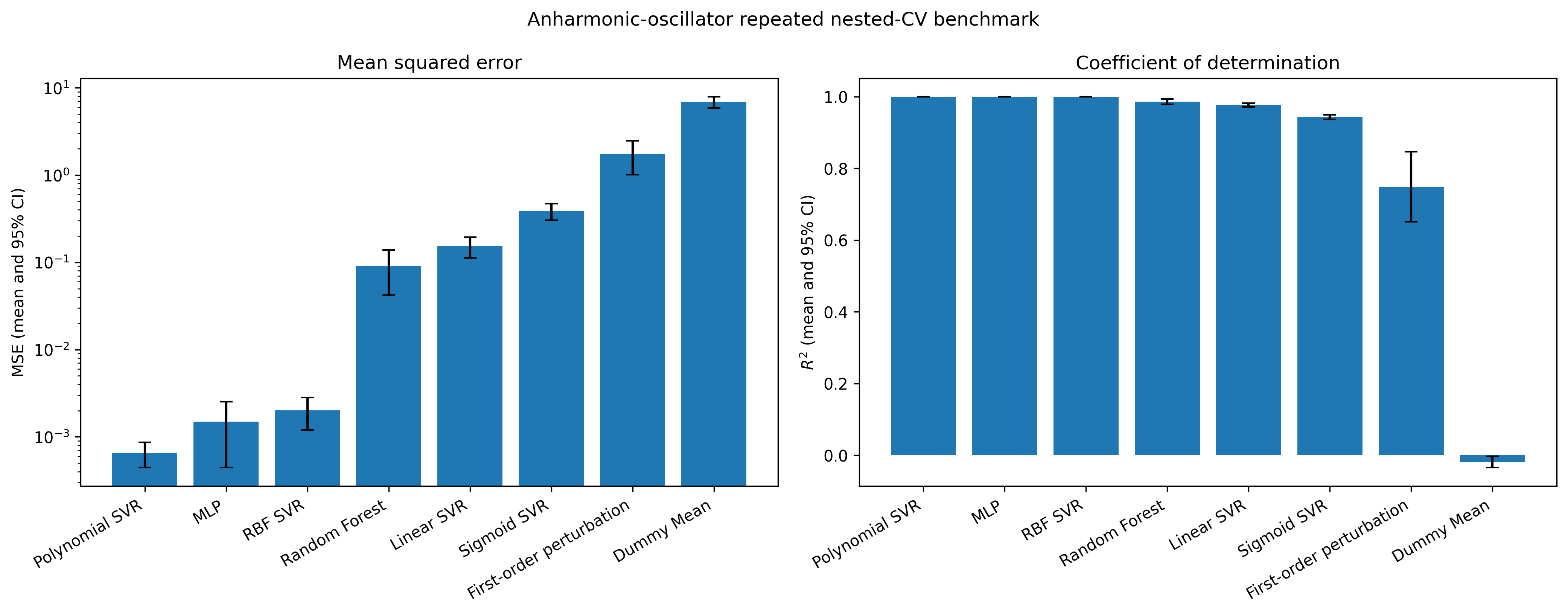}{0.36\textwidth}\hfill
\safeinclude{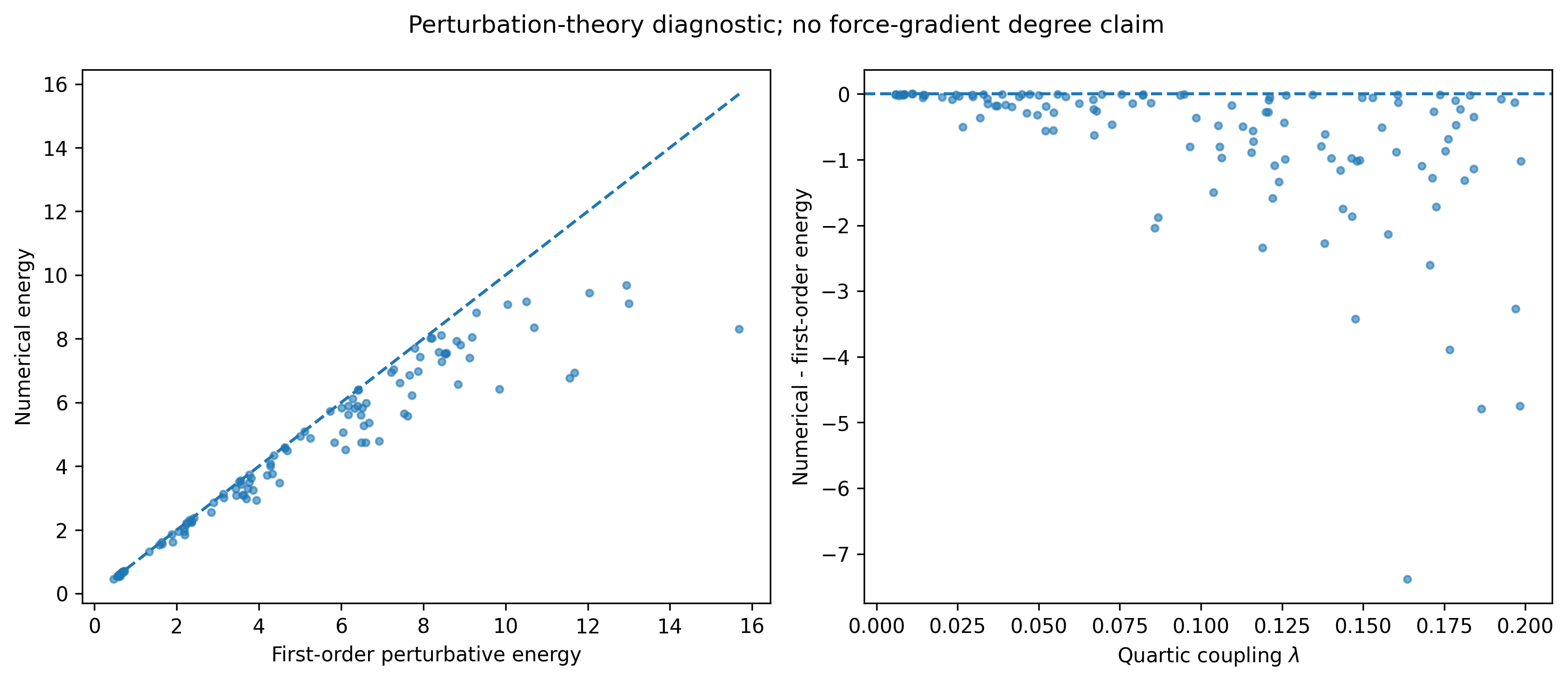}{0.28\textwidth}
\caption{Quartic-oscillator validation and perturbation diagnostic. Left: polynomial-degree ablation under the same validation protocol; degree 3 gives the lowest mean MSE among degrees 1--4 on the sampled domain, while the result is not interpreted as a consequence of the cubic force. Center: repeated nested-CV comparison of the validation-selected polynomial SVR with alternative model families and analytical/baseline predictors. Right: first-order perturbative energy versus the numerical spectrum and the residual $\Delta E_n=E_n^{\mathrm{num}}-E_n^{(1)}$ versus quartic coupling $\lambda$; increasing negative residuals and dispersion at larger $\lambda$ indicate the loss of first-order accuracy.}
\label{fig:anharmonic}
\end{figure*}

Figure~\ref{fig:anharmonic} summarizes what the numerical data show. In the degree-ablation panel, the degree-3 polynomial SVR has the smallest mean cross-validated MSE among the tested degrees $1$--$4$, with degree 4 slightly worse and degrees 1--2 producing larger errors. The corresponding $R^2$ values remain high for the better-performing degrees. Thus degree 3 is an empirical optimum on the sampled $(n,k,\lambda)$ domain; it is not derived from the quartic power of the potential or from the cubic force.

The center panel places this selected polynomial SVR in the same repeated nested-CV benchmark as the alternative model families. Several nonlinear data-driven regressors remain competitive, whereas the first-order perturbative expression is less accurate once the sampled parameters move beyond the weak-coupling regime, and the mean baseline performs poorly. The purpose of this comparison is therefore not to claim that one kernel degree is universally preferred, but to test whether the validation-selected polynomial representation remains competitive against unrelated regressors.

The perturbation diagnostic in the right panel supplies the physical interpretation independently of the kernel-degree choice. Numerical energies follow the first-order prediction closely in the weakly anharmonic part of the sampled domain, whereas the deviation grows as $\lambda$ increases. With the residual defined as
\begin{equation}\label{eq:quartic_residual_diagnostic}
\Delta E_n=E_n^{\mathrm{num}}-E_n^{(1)},
\end{equation}
the increasingly negative and more dispersed residuals at larger quartic coupling show that first-order perturbation theory progressively overestimates part of the numerical spectrum and that higher-order or nonperturbative corrections become relevant. Taken together, the three panels support two separate conclusions: perturbation theory provides a physically interpretable baseline, while the finite-domain polynomial degree is determined by validation rather than by the force gradient.

\subsection{Photonic-crystal transmission}\label{sec:crystals}
A finite one-dimensional Si/SiO$_2$ multilayer is evaluated by a normal-incidence transfer-matrix calculation \cite{Joannopoulos2008}. One-dimensional dielectric stacks are a canonical setting in which multiple reflections and phase accumulation produce Bragg-like stop bands and oscillatory pass-band structure. This well-established theoretical framework motivates the use of a deliberately simple physical generator for the present benchmark: silicon dispersion is modeled by the stated empirical refractive-index relation, absorption is neglected, and silica is treated as nondispersive.


The incident medium, finite layer sequence, and exit medium define the boundary-value problem before regression. Energy is the scalar input and the simulated transmission is the target. The exact Jackson--Chebyshev model, Nystr\"om features, standard SVR kernels, random forest, and MLP are compared under the same validation logic. Rather than by pointwise equality to a Maxwell Green function, the custom kernel is motivated by the spectral morphology of the response: sharp band edges, oscillations, and repeated structure. Its explicit Chebyshev modes and Jackson damping provide a controlled way to represent that structured spectrum; they are not claimed to reconstruct Maxwell boundary operators.

\begin{figure*}[htb]
\centering
\safeinclude{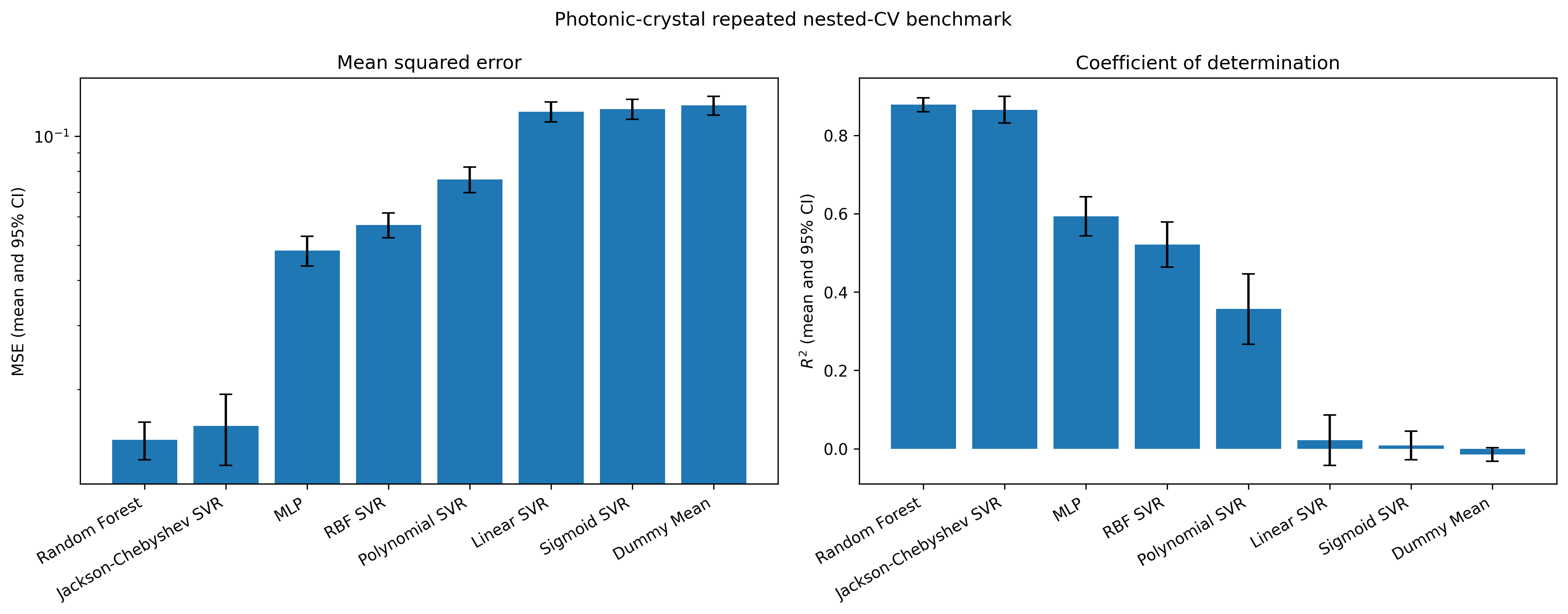}{0.52\textwidth}\hfill
\safeinclude{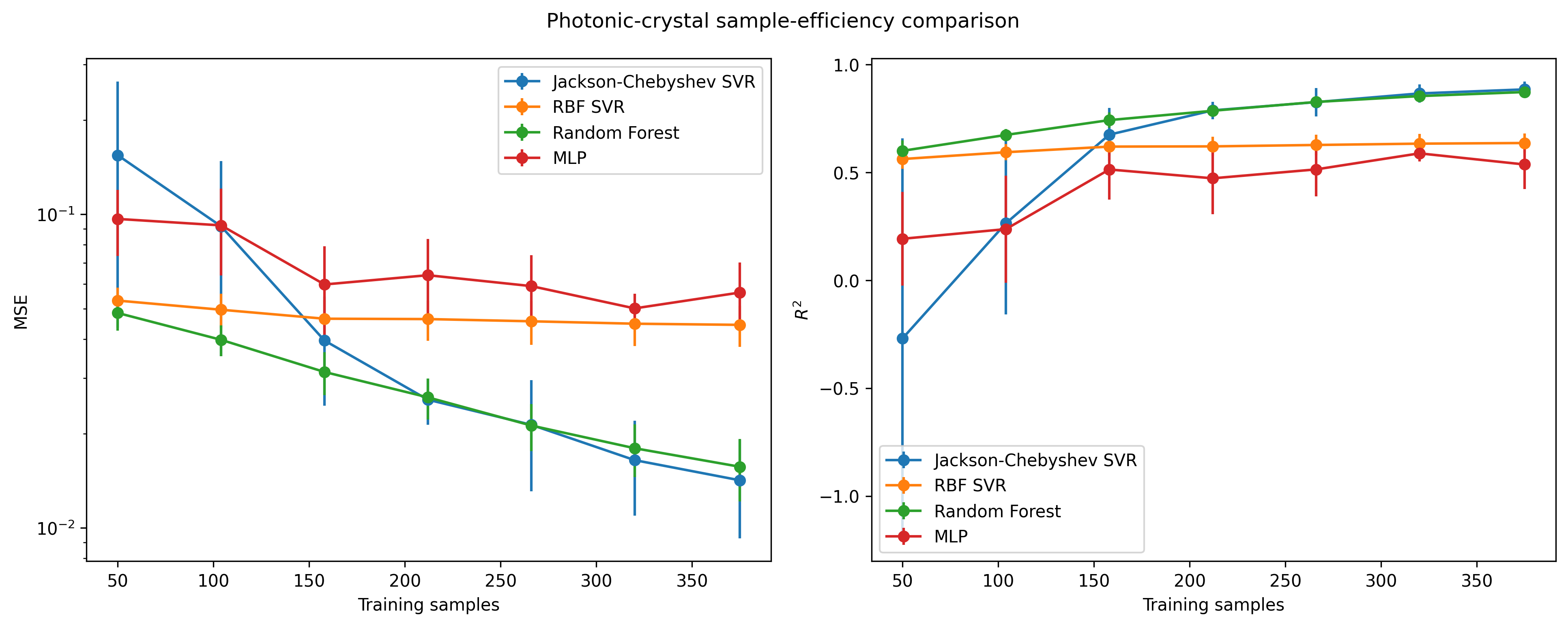}{0.44\textwidth}
\caption{Photonic-crystal benchmark and sample-efficiency analysis. Left: repeated nested-CV mean MSE (lower is better) and $R^2$ (higher is better), with 95\% confidence intervals, for Jackson-Chebyshev SVR, standard SVR kernels, random forest, MLP, and the mean baseline. Random forest and Jackson-Chebyshev SVR form the leading group, while the standard smooth kernels are less well matched to the oscillatory response. Right: MSE and $R^2$ learning curves versus the number of training samples. Jackson-Chebyshev SVR improves markedly with sample density and approaches the random-forest curve at the larger sample sizes, whereas RBF SVR plateaus earlier. 
}
\label{fig:photonic_main}
\end{figure*}

Figure~\ref{fig:photonic_main} should be read as two complementary tests. In the repeated nested-CV benchmark, random forest and Jackson--Chebyshev SVR form the leading group, whereas RBF, polynomial, linear, and sigmoid kernels progressively lose accuracy on the oscillatory transmission task. The learning curves then show how this ranking evolves with sample size: the Jackson--Chebyshev model improves strongly as additional spectral samples are supplied and approaches the random-forest level, while the RBF curve saturates earlier. Thus the figure shows that an explicit oscillatory spectral basis is a competitive and increasingly effective inductive bias as the structured spectrum is more densely sampled.

\begin{figure*}[htb]
\centering
\safeinclude{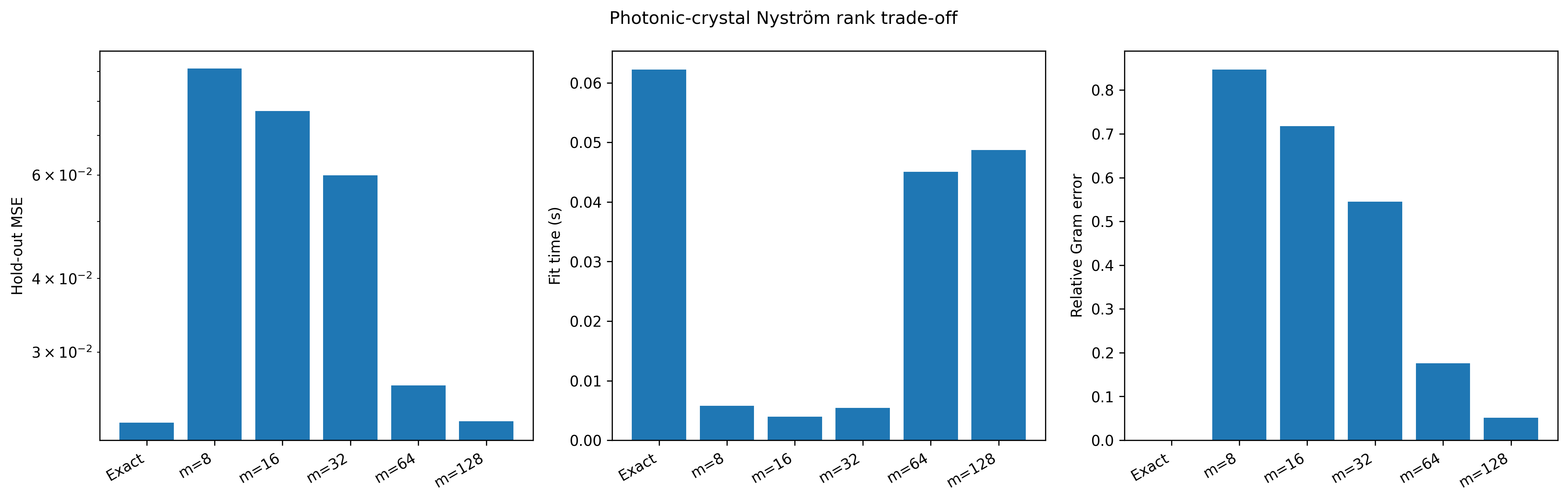}{0.55\textwidth}\hfill
\safeinclude{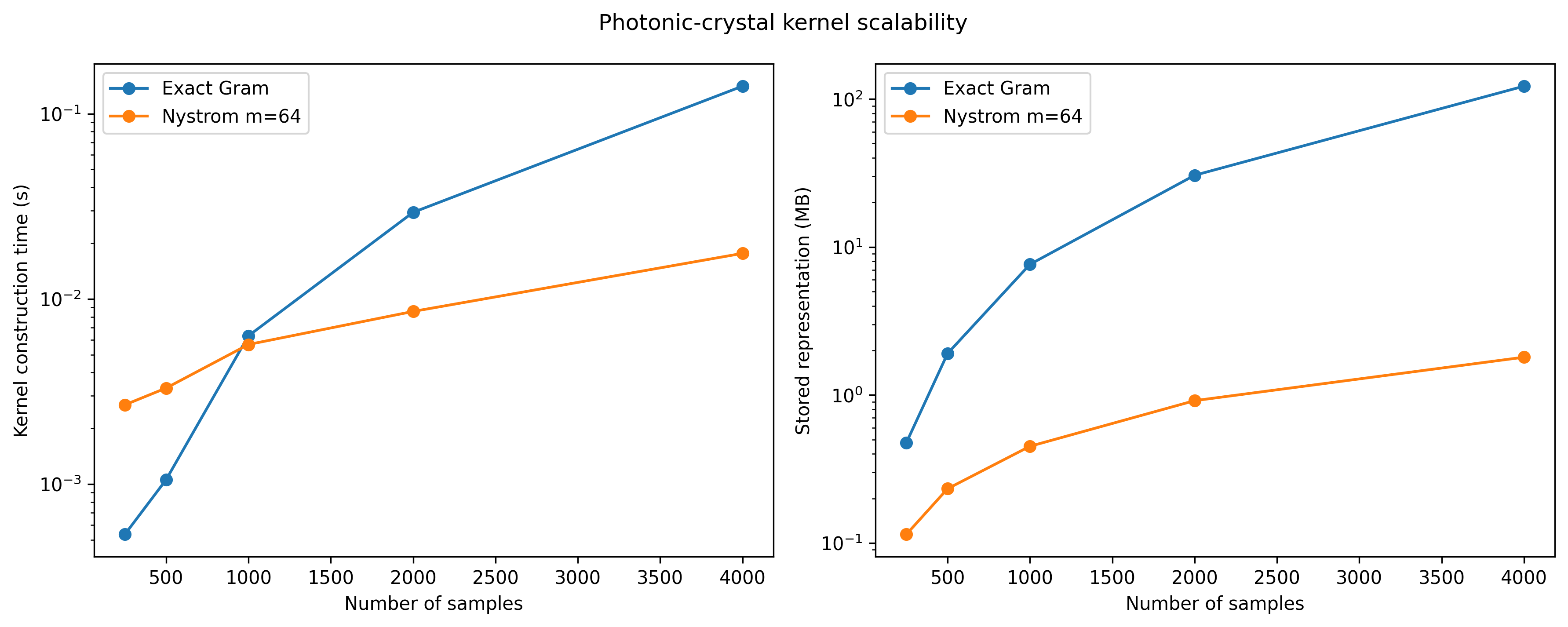}{0.41\textwidth}
\caption{Photonic-crystal Nystr\"om rank--accuracy--cost analysis. Here $m$ is the number of Nystr\"om landmark samples (hence $\operatorname{rank}(\widetilde K)\le m$), not a physical system parameter. Left block: hold-out MSE, fit time, and relative Gram-matrix error for the exact kernel and $m=8,16,32,64,128$. Small $m$ is computationally cheap but produces a poor Gram approximation and larger predictive error; increasing $m$ systematically reduces the Gram error and recovers the exact-kernel predictive level, at increasing cost. Right block: exact-Gram versus fixed-$m=64$ Nystr\"om kernel-construction time and stored representation size as the sample count $n$ grows. The low-rank representation changes the storage scaling from a full $n\times n$ matrix toward an $n\times m$ representation and becomes increasingly advantageous at larger $n$. No spectral clipping is used as a PSD-repair step.}
\label{fig:photonic_scaling}
\end{figure*}

Before interpreting the Nystr\"om plots, we note that $m$ is a numerical approximation parameter, not a physical material quantity: it is the number of landmark training samples in $L=\{\ell_a\}_{a=1}^{m}$, and the approximation $\widetilde K=ZZ^\top$ has rank at most $m$. Figure~\ref{fig:photonic_scaling} therefore asks how much predictive and Gram-matrix fidelity is retained when the full $n\times n$ representation is compressed to an $m$-landmark representation.

The Nystr\"om rank sweep in Fig.~\ref{fig:photonic_scaling} makes the tradeoff quantitative. At very small $m$, the approximation is fast but the relative Gram error is large and the hold-out MSE is correspondingly worse. As $m$ increases, the approximation approaches the exact-kernel reference; by the largest landmark counts shown, the predictive error is close to the exact result and the Gram error is much smaller, although the fit cost has risen. The accompanying scalability panel holds $m=64$ fixed while increasing $n$ and shows why the approximation matters operationally: the stored representation grows far more slowly than the exact dense Gram matrix, and the kernel-construction time becomes lower once the sample size is sufficiently large. 

\subsection{Boundary-conditioned Fibonacci-chain transmission}\label{sec:quasi}
A finite Fibonacci tight-binding chain provides a quasiperiodic, resonance-rich spectral-regression test. Fibonacci chains are a standard minimal setting for studying deterministic aperiodic order: the nonperiodic sequence produces a fragmented, highly structured spectral response rather than the simple bands of a periodic chain. To turn the finite sequence into an open transport problem, the device is coupled to left and right semi-infinite one-dimensional leads. For retarded lead self-energies $\Sigma_L^R$ and $\Sigma_R^R$,
\begin{equation}
\begin{split}
G^R(E)=\big[&(E+i\eta)I-H\\
&-\Sigma_L^R(E)-\Sigma_R^R(E)\big]^{-1}.
\end{split}
\end{equation}
and the single-channel Landauer transmission is
\begin{equation}
\begin{split}
T(E)&=\Gamma_L(E)\Gamma_R(E)|G^R_{1N}(E)|^2,\\
\Gamma_{L/R}&=-2\,\mathrm{Im}\,\Sigma_{L/R}^R.
\end{split}
\end{equation}
consistent with standard mesoscopic transport \cite{Datta1995}. This example therefore has the most direct Green-function role among the five benchmarks: the lead-dressed physical resolvent is used explicitly in the data generator to produce $T(E)$, and the self-energies encode the open boundaries. The Jackson--Chebyshev regression kernel acts only on the sampled energy variable and is not identified with that resolvent. Its role is instead to supply an inspectable spectral interpolation prior for the resonance-rich function generated by the open quantum system.

\begin{figure*}[htb]
\centering
\safeinclude{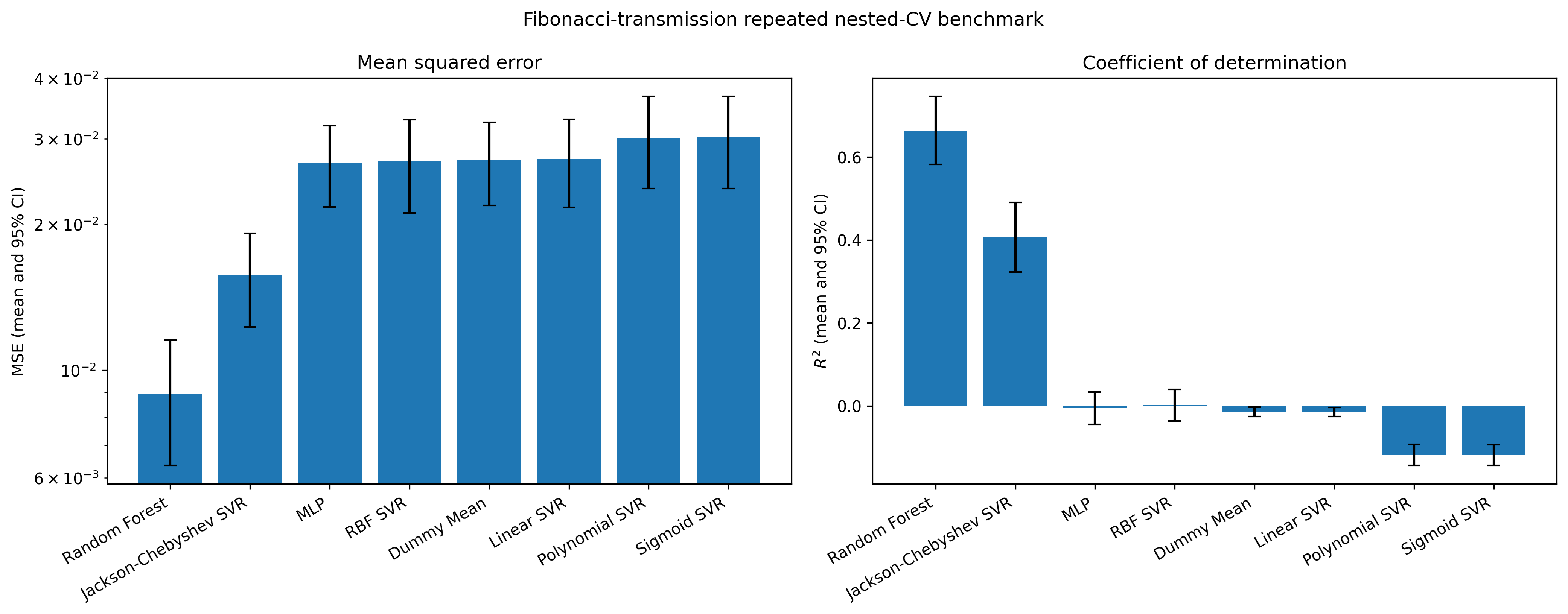}{0.48\textwidth}\hfill
\safeinclude{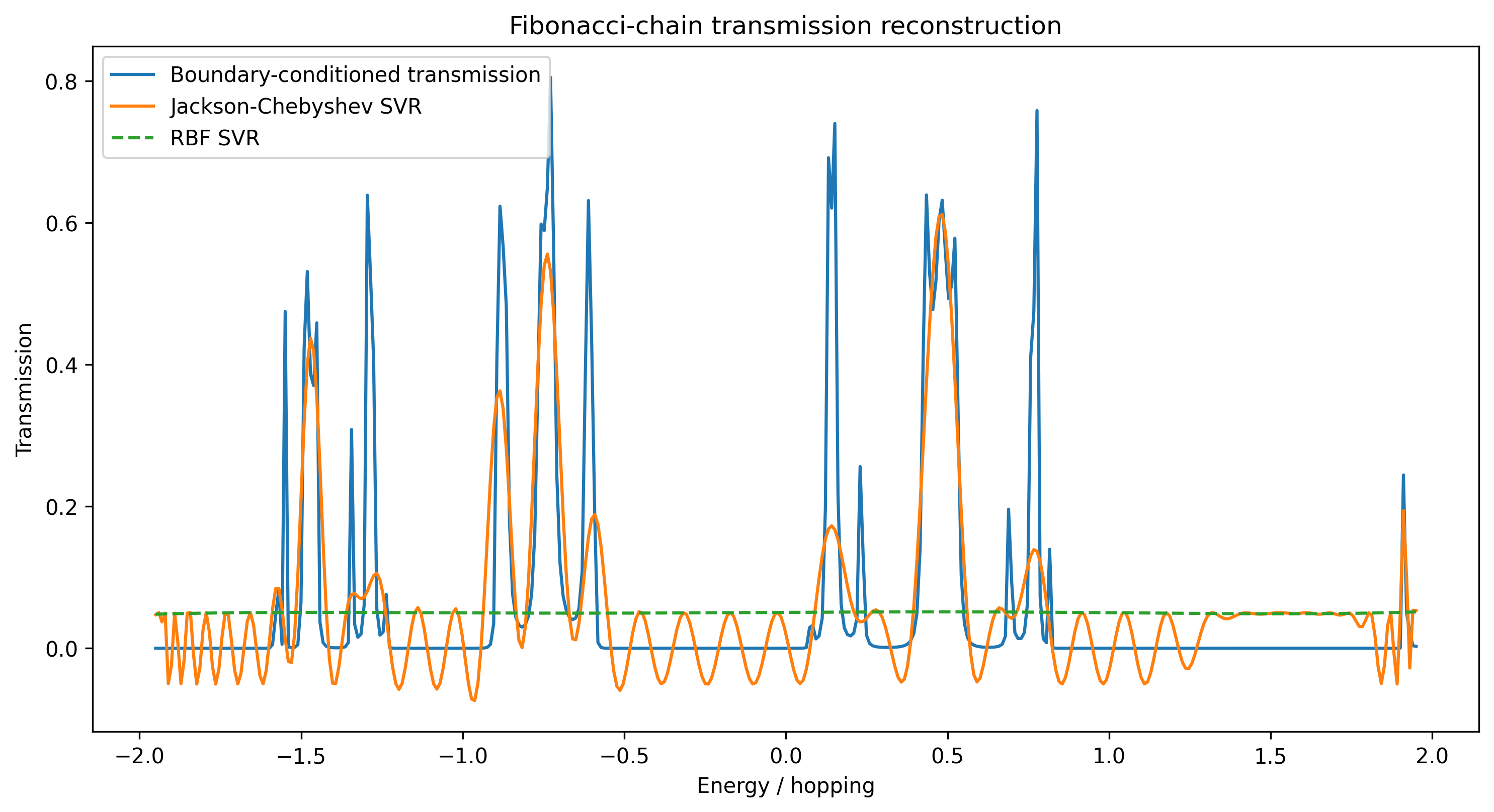}{0.48\textwidth}
\caption{Boundary-conditioned Fibonacci-chain benchmark and spectral reconstruction. All curves and model scores refer to the same boundary-conditioned target generated with the semi-infinite-lead self-energies; this is not a boundary/no-boundary comparison. Left: repeated nested-CV MSE and $R^2$ for the regressors. Random forest gives the best aggregate score, Jackson--Chebyshev is the strongest SVR-type model, and RBF SVR is close to a mean-like fit. Right: the blue curve is the physical transmission $T(E)$ generated from the lead-dressed Green function, the orange curve is the Jackson--Chebyshev SVR reconstruction, and the dashed green curve is the RBF-SVR reconstruction. The flat RBF prediction illustrates over-smoothing of narrow resonances, while Jackson--Chebyshev captures more of the resonance envelope but not every sharp peak. Small negative Jackson--Chebyshev excursions are artifacts of unconstrained regression rather than physical negative transmission.}
\label{fig:quasi_main}
\end{figure*}

In Figure~\ref{fig:quasi_main}, all targets and all regressors use the same boundary-conditioned transmission data generated from Eqs.~(34)--(35). The left panel compares regression models on that common target: random forest has the best aggregate nested-CV score, Jackson--Chebyshev SVR is the strongest SVR-type model, and RBF SVR performs close to a nearly mean-like prediction. The right panel explains this difference spectrally. The blue curve is the physical transmission generated from the lead-dressed Green function; the orange curve is the Jackson--Chebyshev reconstruction; and the dashed green curve is the RBF-SVR reconstruction. The nearly horizontal RBF curve is therefore an underfitting/over-smoothing outcome for narrow, sparsely distributed resonances, not the result of omitting boundary conditions. Jackson--Chebyshev captures several broad resonance locations and envelopes but smooths some sharp peaks. Its small negative excursions are also nonphysical ($T$ should lie in $[0,1]$) and reveal a limitation of unconstrained regression, motivating future target transformations or output constraints.

\begin{figure*}[htb]
\centering
\safeinclude{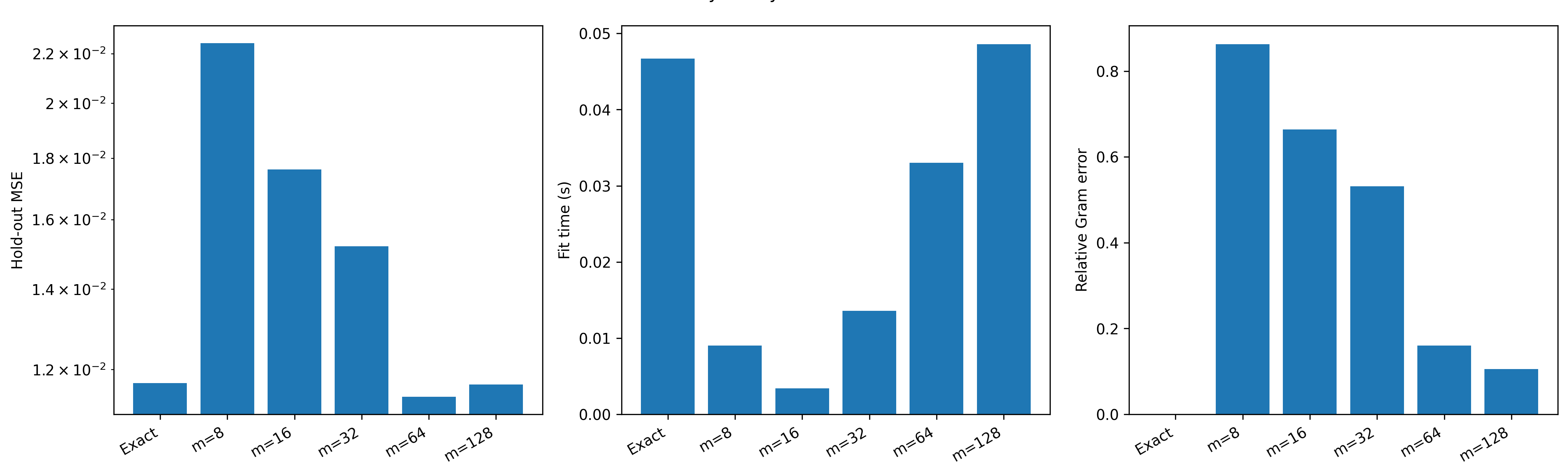}{0.48\textwidth}\hfill
\safeinclude{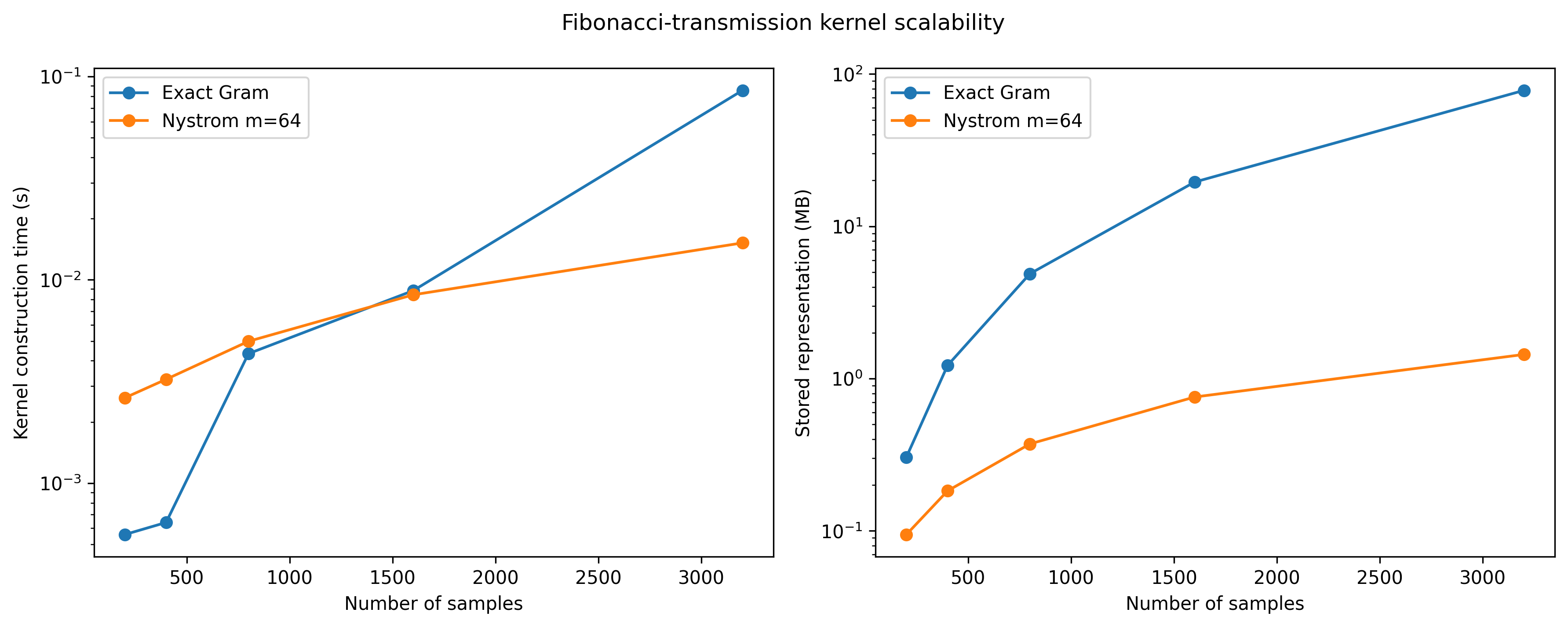}{0.48\textwidth}
\caption{Fibonacci-chain Nystr\"om rank--accuracy--cost analysis. Here $m$ is the number of landmark samples and $\operatorname{rank}(\widetilde K)\le m$. Left block: hold-out MSE, fit time, and relative Gram-matrix error for the exact representation and increasing $m$. The smallest ranks are fastest but least faithful; increasing $m$ sharply reduces the Gram error and brings the predictive MSE toward the exact-kernel level, with a corresponding increase in fit time. Right block: exact-Gram and fixed-$m=64$ Nystr\"om kernel-construction time and stored representation size versus the number of samples $n$. The approximation is therefore evaluated jointly for predictive fidelity, matrix fidelity, time, and memory rather than by runtime alone.}
\label{fig:quasi_scaling}
\end{figure*}

As in the photonic test, the symbol $m$ in the Fibonacci Nystr\"om analysis denotes the number of landmark samples and hence an upper bound on the low-rank representation. Figure~\ref{fig:quasi_scaling} asks whether the resonance-rich data can be compressed in kernel space without sacrificing the predictive behavior of the exact Jackson--Chebyshev kernel.

Figure~\ref{fig:quasi_scaling} provides the same rank-accuracy-cost check for the Fibonacci-chain data. The $m=8$ and $m=16$ approximations are cheap but have large relative Gram errors and noticeably worse hold-out MSE. Increasing the landmark count substantially reduces the Gram error. Around the larger ranks the predictive MSE approaches the exact-kernel reference, although the fitting cost rises. In the fixed $m=64$ scalability panel, the Nystr\"om representation avoids storage of the full $n\times n$ matrix and becomes increasingly advantageous as $n$ grows. 


\section{Conclusion}\label{sec:conclusion}
We developed a physics-informed kernel-design framework for finite-data regression of scalar or low-dimensional physical observables. The Green-function connection is deliberately hierarchical rather than universal. It is an exact functional identity only in special cases, a source of operator/spectral structure more generally, and an explicit part of the physical data generator in the open Fibonacci transport example. A Mercer kernel is not a physical resolvent, SVR dual coefficients are not physical source amplitudes, and boundary conditions remain part of the physical operator or simulator unless a boundary-adapted kernel is constructed explicitly.

The Jackson--Chebyshev construction provides an explicit spectral feature basis with nonnegative Jackson weights. Consequently, its Gram matrix is PSD by construction, so the cubic full-matrix spectral-clipping step is unnecessary. Exact dense-kernel SVR still has quadratic storage requirements; the Nystr\"om representation supplies a lower-rank alternative whose approximation error, predictive error, time, and memory are measured directly.

The experiments should be interpreted in terms of inductive bias rather than universal predictive dominance. The copper and local-dispersion examples retain concise physical derivations that motivate standard-kernel priors. For the quartic oscillator, perturbation theory supplies an analytical baseline, while polynomial degree is selected empirically by nested validation rather than inferred from the cubic force. In the transmission examples, boundary conditions are explicit in the physical generators, and the custom kernel supplies a separate, inspectable spectral interpolation prior.

Random forest or MLPs may outperform the proposed method on individual finite data sets. The value of the present approach is instead its explicit basis, damping profile, kernel--target alignment, and support-vector diagnostics. 

\section*{Acknowledgments}
The authors thank Professor Chong-Der Hu of National Taiwan University and Assistant Professor Tsung-Wei Chiang of National Chung Cheng University for valuable discussions. Renata Wong acknowledges support from National Science and Technology Council grant No.~NSTC 114-2112-M-182-002-MY3 and Chang Gung Memorial Hospital grant No.~BMRPL94.

\section*{Code and data availability}
The code is available at \url{https://github.com/kuonanhong/SVM-kernels-as-quantum-propagators}. Materials Project credentials are read from a private environment variable or local untracked key file and are not committed to the public repository.

\bibliographystyle{plain}

\begin{thebibliography}{99}
\bibitem{Cortes1995} C.~Cortes and V.~Vapnik, ``Support-vector networks,'' \emph{Machine Learning}, vol.~20, no.~3, pp.~273--297, 1995.
\bibitem{Vapnik1999} V.~N.~Vapnik, \emph{The Nature of Statistical Learning Theory}, Springer, 1999.
\bibitem{Scholkopf2002} B.~Sch\"olkopf and A.~J.~Smola, \emph{Learning with Kernels: Support Vector Machines, Regularization, Optimization, and Beyond}, MIT Press, 2002.
\bibitem{Feynman1965} R.~P.~Feynman and A.~R.~Hibbs, \emph{Quantum Mechanics and Path Integrals}, McGraw-Hill, 1965.
\bibitem{Deeter1974} C.~R.~Deeter and J.~M.~Gray, ``The discrete Green's function and the discrete kernel function,'' \emph{Discrete Mathematics}, vol.~10, no.~1, pp.~29--42, 1974.
\bibitem{Davies1987} E.~B.~Davies, ``The equivalence of certain heat kernel and Green function bounds,'' \emph{Journal of Functional Analysis}, vol.~71, no.~1, pp.~88--103, 1987.
\bibitem{Fasshauer2011} G.~E.~Fasshauer, ``Green's Functions: Taking Another Look at Kernel Approximation, Radial Basis Functions, and Splines,'' \emph{Springer Proceedings in Mathematics}, vol.~13, pp.~37--63, 2011.
\bibitem{Dean2021} D.~S.~Dean, P.~Le Doussal, S.~N.~Majumdar, G.~Schehr, and N.~R.~Smith, ``Kernels for non interacting fermions via a Green's function approach with applications to step potentials,'' \emph{Journal of Physics A: Mathematical and Theoretical}, vol.~54, 084001, 2021.
\bibitem{Gin2021} C.~R.~Gin, D.~E.~Shea, S.~L.~Brunton, and J.~N.~Kutz, ``DeepGreen: deep learning of Green's functions for nonlinear boundary value problems,'' \emph{Scientific Reports}, vol.~11, 21614, 2021.
\bibitem{Li2020} Z.~Li, N.~Kovachki, K.~Azizzadenesheli, B.~Liu, K.~Bhattacharya, A.~Stuart, and A.~Anandkumar, ``Neural operator: Graph kernel network for partial differential equations,'' \emph{arXiv:2003.03485}, 2020.
\bibitem{Vapnik1998} V.~Vapnik, \emph{Statistical Learning Theory}, Wiley, 1998.
\bibitem{KimeldorfWahba1971} G.~S.~Kimeldorf and G.~Wahba, ``Some results on Tchebycheff spline functions,'' \emph{Journal of Mathematical Analysis and Applications}, vol.~33, no.~1, pp.~82--95, 1971.
\bibitem{Mercer1909} J.~Mercer, ``Functions of positive and negative type and their connection with the theory of integral equations,'' \emph{Philosophical Transactions of the Royal Society A}, vol.~209, pp.~415--446, 1909.
\bibitem{CourantHilbert1962} R.~Courant and D.~Hilbert, \emph{Methods of Mathematical Physics}, vol.~2, Wiley-Interscience, 1962.
\bibitem{SupportVectorIndefinite} I.~Alabdulmohsin, X.~Gao, and X.~Zhang, ``Support Vector Machines with Indefinite Kernels,'' in \emph{Proceedings of the 31st International Conference on Machine Learning (ICML)}, pp.~32--47, 2015.
\bibitem{TrainingSVMIndefinite} J.~Chen and J.~Ye, ``Training SVM with Indefinite Kernels,'' in \emph{Proceedings of the 25th International Conference on Machine Learning (ICML)}, pp.~232--239, 2008.
\bibitem{SolvingIndefiniteSVM} H.-M.~Xu, H.~Xue, X.-H.~Chen, and Y.-Y.~Wang, ``Solving Indefinite Kernel Support Vector Machine with Difference of Convex Functions Programming,'' in \emph{Proceedings of the 31st AAAI Conference on Artificial Intelligence}, pp.~2782--2788, 2017.
\bibitem{SupportVectorIndefiniteKrein} C.~Loosli, G.~Canu, and C.~S.~Ong, ``Learning SVMs with indefinite kernels,'' in \emph{Proceedings of ICANN}, vol.~1, pp.~442--451, 2012.
\bibitem{AnalysisIndefiniteKernels} Y.~Ying, C.~Campbell, and M.~Girolami, ``Analysis of SVM with Indefinite Kernels,'' in \emph{Proceedings of ICML}, pp.~687--694, 2012.
\bibitem{IndefiniteKernelsPCA} X.~Huang, A.~Maier, J.~Hornegger, and J.~A.~K.~Suykens, ``Indefinite kernels in least squares support vector machines and principal component analysis,'' \emph{Applied and Computational Harmonic Analysis}, vol.~43, no.~2, pp.~162--172, 2017.
\bibitem{SVMIndefiniteClassification} R.~Luss and A.~d'Aspremont, ``Support vector machine classification with indefinite kernels,'' in \emph{Proceedings of ICML}, pp.~321--328, 2007.
\bibitem{SVMWithIndefiniteKernels} I.~Alabdulmohsin, X.~Gao, and X.~Zhang, ``Support vector machines with indefinite kernels,'' \emph{JMLR Workshop and Conference Proceedings}, vol.~39, pp.~32--47, 2014.
\bibitem{SigmoidSMO} H.-T.~Lin and C.-J.~Lin, ``A Study on Sigmoid Kernels for SVM and the Training of Non-PSD Kernels by SMO-Type Methods,'' \emph{Neural Computation}, vol.~16, no.~5, pp.~1071--1090, 2004.
\bibitem{platt} J.~C.~Platt, ``Fast training of support vector machines using sequential minimal optimization,'' in B.~Sch\"olkopf, C.~J.~C.~Burges, and A.~J.~Smola (eds.), \emph{Advances in Kernel Methods: Support Vector Machines}, MIT Press, pp.~185--208, 1998.
\bibitem{Silver1994} R.~N.~Silver and H.~R\"oder, ``Calculation of densities of states and spectral functions by Chebyshev recursion and maximum entropy,'' \emph{Physical Review E}, vol.~56, no.~4, pp.~4822--4829, 1997.
\bibitem{Weisse2006} A.~Wei\ss e, G.~Wellein, A.~Alvermann, and H.~Fehske, ``The kernel polynomial method,'' \emph{Reviews of Modern Physics}, vol.~78, no.~1, pp.~275--306, 2006.
\bibitem{JacobiKPM2025} I.~O.~Raikov and Y.~M.~Beltukov, ``The kernel polynomial method based on Jacobi polynomials,'' \emph{Applied Mathematics and Computation}, vol.~490, 129207, 2025, doi:10.1016/j.amc.2024.129207.
\bibitem{materials_project} A.~Jain \emph{et al.}, ``Commentary: The Materials Project: A materials genome approach to accelerating materials innovation,'' \emph{APL Materials}, vol.~1, no.~1, 011002, 2013.
\bibitem{mp2}Horton, M.K., Huck, P., Yang, R.X. \emph{et al.}, ``Accelerated data-driven materials science with the Materials Project,'' \emph{Nat. Mater.} 24, 1522–1532 (2025). https://doi.org/10.1038/s41563-025-02272-0




\bibitem{Drucker1997} H.~Drucker, C.~J.~C.~Burges, L.~Kaufman, A.~Smola, and V.~Vapnik, ``Support Vector Regression Machines,'' in \emph{Advances in Neural Information Processing Systems}, vol.~9, MIT Press, 1997.
\bibitem{Mahan1990} G.~D.~Mahan, \emph{Many-Particle Physics}, 2nd ed., Plenum Press, 1990.
\bibitem{Genton2001} M.~G.~Genton, ``Classes of Kernels for Machine Learning: A Statistics Perspective,'' \emph{Journal of Machine Learning Research}, vol.~2, pp.~299--312, 2001.
\bibitem{Sakurai2014} J.~J.~Sakurai and J.~Napolitano, \emph{Modern Quantum Mechanics}, 2nd ed., Pearson, 2014.
\bibitem{Landau1977} L.~D.~Landau and E.~M.~Lifshitz, \emph{Quantum Mechanics: Non-Relativistic Theory}, 3rd ed., Pergamon Press, 1977.
\bibitem{Joannopoulos2008} J.~D.~Joannopoulos, S.~G.~Johnson, J.~N.~Winn, and R.~D.~Meade, \emph{Photonic Crystals: Molding the Flow of Light}, 2nd ed., Princeton University Press, 2008.
\bibitem{graphene} S.~R.~Power and M.~S.~Ferreira, ``Electronic structure of graphene beyond the linear dispersion regime,'' \emph{Physical Review B}, vol.~83, 155432, 2011.
\bibitem{neto} A.~H.~Castro Neto, F.~Guinea, N.~M.~R.~Peres, K.~S.~Novoselov, and A.~K.~Geim, ``The electronic properties of graphene,'' \emph{Reviews of Modern Physics}, vol.~81, 109, 2009.
\bibitem{lanczos} C.~Lanczos, ``An iteration method for the solution of the eigenvalue problem of linear differential and integral operators,'' \emph{Journal of Research of the National Bureau of Standards}, vol.~45, no.~4, pp.~255--282, 1950.
\bibitem{nystrom} E.~J.~Nystr\"om, ``\"Uber die praktische Aufl\"osung von Integralgleichungen mit Anwendungen auf Randwertaufgaben,'' \emph{Acta Mathematica}, vol.~54, no.~1, pp.~185--204, 1930.

\bibitem{Raissi2019} M.~Raissi, P.~Perdikaris, and G.~E.~Karniadakis, ``Physics-informed neural networks: A deep learning framework for solving forward and inverse problems involving nonlinear partial differential equations,'' \emph{Journal of Computational Physics}, vol.~378, pp.~686--707, 2019.
\bibitem{Lu2021} L.~Lu, P.~Jin, G.~Pang, Z.~Zhang, and G.~E.~Karniadakis, ``Learning nonlinear operators via DeepONet based on the universal approximation theorem of operators,'' \emph{Nature Machine Intelligence}, vol.~3, pp.~218--229, 2021.
\bibitem{LiFNO2021} Z.~Li, N.~Kovachki, K.~Azizzadenesheli, B.~Liu, K.~Bhattacharya, A.~Stuart, and A.~Anandkumar, ``Fourier neural operator for parametric partial differential equations,'' in \emph{International Conference on Learning Representations}, 2021.
\bibitem{WilliamsSeeger2001} C.~K.~I.~Williams and M.~Seeger, ``Using the Nystr\"om method to speed up kernel machines,'' in \emph{Advances in Neural Information Processing Systems 13}, pp.~682--688, 2001.
\bibitem{Cristianini2002} N.~Cristianini, J.~Shawe-Taylor, A.~Elisseeff, and J.~S.~Kandola, ``On kernel-target alignment,'' in \emph{Advances in Neural Information Processing Systems 14}, pp.~367--373, 2002.
\bibitem{Datta1995} S.~Datta, \emph{Electronic Transport in Mesoscopic Systems}, Cambridge University Press, 1995.
\bibitem{LorinYang2024} E.~Lorin and X.~Yang, ``Quasi-optimal domain decomposition method for neural network-based computation of the time-dependent Schr\"odinger equation,'' \emph{Computer Physics Communications}, vol.~299, 109129, 2024, doi:10.1016/j.cpc.2024.109129.
\bibitem{Gamper2024} J.~Gamper, H.~G.~Gallmetzer, A.~K.~H.~Weiss, and T.~S.~Hofer, ``A general strategy for improving the performance of PINNs -- Analytical gradients and advanced optimizers in the NeuralSchr\"odinger framework,'' \emph{Artificial Intelligence Chemistry}, vol.~2, no.~1, 100047, 2024, doi:10.1016/j.aichem.2024.100047.
\bibitem{Ullah2024} A.~Ullah, Y.~Huang, M.~Yang, and P.~O.~Dral, ``Physics-informed neural networks and beyond: enforcing physical constraints in quantum dissipative dynamics,'' \emph{Digital Discovery}, vol.~3, pp.~2052--2060, 2024, doi:10.1039/D4DD00153B.
\bibitem{Kalitenko2025} A.~M.~Kalitenko and P.~I.~Pronin, ``Physics-informed neural network for quantum systems with time-dependent boundary conditions,'' \emph{Physics Letters A}, vol.~560, 130928, 2025, doi:10.1016/j.physleta.2025.130928.
\bibitem{Lanthaler2025} S.~Lanthaler, Z.~Li, and A.~M.~Stuart, ``Nonlocality and Nonlinearity Implies Universality in Operator Learning,'' \emph{Constructive Approximation}, vol.~62, pp.~261--303, 2025, doi:10.1007/s00365-025-09718-3.
\end{thebibliography}

\end{document}